\documentclass[epj]{svjour}
\usepackage{amsmath,amssymb}
\usepackage{graphicx}
\usepackage{enumerate}
\usepackage{changebar,newcent,epsfig,color,url}

\begin{document}
\title{The Geometry of Chaotic Dynamics -- A Complex Network Perspective}
\author{Reik V. Donner\inst{1} \and Jobst Heitzig\inst{1} \and Jonathan F. Donges\inst{1,2} \and Yong Zou\inst{1} \and Norbert Marwan\inst{1} \and J\"urgen Kurths\inst{1,2}}
\institute{Potsdam Institute for Climate Impact Research, P.O. Box 60\,12\,03, 14412 Potsdam, Germany \and Department of Physics, Humboldt University Berlin, Newtonstr.~15, 12489 Berlin, Germany}
\offprints{reik.donner@pik-potsdam.de}

\date{\today}

\abstract{Recently, several complex network approaches to time series analysis have been developed and applied to study a wide range of model systems as well as real-world data, e.g., geophysical or financial time series. Among these techniques, recurrence-based concepts and prominently $\varepsilon$-recurrence networks, most faithfully represent the geometrical fine structure of the attractors underlying chaotic (and less interestingly non-chaotic) time series. In this paper we demonstrate that the well known graph theoretical properties local clustering coefficient and global (network) transitivity can meaningfully be exploited to define two new local and two new global measures of dimension in phase space: local upper and lower clustering dimension as well as global upper and lower transitivity dimension. Rigorous analytical as well as numerical results for self-similar sets and simple chaotic model systems suggest that these measures are well-behaved in most non-pathological situations and that they can be estimated reasonably well using $\varepsilon$-recurrence networks constructed from relatively short time series. Moreover, we study the relationship between clustering and transitivity dimensions on the one hand, and traditional measures like pointwise dimension or local Lyapunov dimension on the other hand. We also provide further evidence that the local clustering coefficients, or equivalently the local clustering dimensions, are useful for identifying unstable periodic orbits and other dynamically invariant objects from time series. Our results demonstrate that $\varepsilon$-recurrence networks exhibit an important link between dynamical systems and graph theory.}

\maketitle

\section{Introduction}\label{sec1}

Recurrence of previous states is a key property of dynamical systems~\cite{marwan2007}. In mathematical terms, one speaks of a recurrence if at time $t_j$, a trajectory of a complex system returns into the dynamical neighbourhood of a previous state $x_i=x(t_i)$ ($t_i<t_j$). Such a neighbourhood can be defined by either considering the $k$-nearest neighbours of $x_i$ (fixed mass of the neighbourhood) or in terms of an $\varepsilon$-ball $B_\varepsilon(x_i)$ centered at $x_i$ (fixed volume of the neighbourhood). In the latter case, one defines the binary \textit{recurrence matrix}
\begin{equation}
R_{ij}(\varepsilon)=\Theta(\varepsilon-\|x_i-x_j\|)
\end{equation}
\noindent
for a trajectory sampled at a fixed number of points $t_i$ in time, where $\Theta(\cdot)$ is the Heaviside function and $\|\cdot\|$ some norm (e.g., maximum or Euclidean norm) in the metric space including the considered attractor. From this definition, it is evident that a recurrence directly corresponds to $R_{ij}=1$. For the sake of clarity, we will more specifically speak about an $\varepsilon$-recurrence in the following to distinguish this definition from the alternative $k$-nearest neighbour based definition. The recurrence matrix can be easily visualised in terms of \textit{recurrence plots} (RPs)~\cite{marwan2007,Eckmann1987,marwan2008epjst}, the appearance of which allows for a simple graphical discrimination between qualitatively different types of dynamics. 

Since the poineering work of Poincar\'e~\cite{poincare1890} on the three-body problem, it has been more and more recognised that recurrences are important for understanding the overall dynamics~\cite{marwan2008epjst}, i.e., they encode all relevant information about the behaviour of a stationary system. Even more, it has been proven recently that the temporal pattern of recurrences allows us to reconstruct the rank-order of any scalar time series (i.e., a unique representation of the original trajectory up to a scaling by some monotonous function)~\cite{Thiel2004PLA,Hirata2008,Robinson2009}. As a consequence, many well established dynamical invariants (such as correlation entropy, correlation dimension, or 2nd-order mutual information) can be estimated from RPs~\cite{thiel2004a}. Moreover, the statistical analysis of line structures in RPs known as \textit{recurrence quantification analysis (RQA)} allows the definition of a large variety of additional measures of complexity, which correspond to time intervals with similar consecutive states (vertical/horizontal lines) or time evolution (diagonal lines off the main diagonal), respectively. Since these measures are easily calculable, RQA has found numerous applications in the last two decades~\cite{marwan2007}.

Besides the phenomenon of recurrence, another appealing paradigm that has attracted considerable interest over the last years are networks with complex topology, called shortly complex networks. In the most simple case (to which we will restrict our attention in this work), a network can be mathematically described as simple graph $G=(V,E)$, where $V=\{1,\dots,N\}$ is the set of vertices with $|V|=N$, and $E\subseteq V\times V$ is the set of edges between pairs of vertices. Note that we do neither consider multiple edges between two vertices, nor hyperedges connecting more than two vertices with each other, i.e. $|E|\leq N(N-1)/2$. Furthermore, we will restrict our attention to unweighted and undirected networks. In this case, the whole network connectivity is completely described by the symmetric \textit{adjacency matrix}
\begin{equation}
A_{ij}=A_{ji}=\left\{ \begin{array}{ll} 1, & \quad (i,j)\in E \\ 0, & \quad (i,j)\not\in E \end{array} \right. .
\end{equation}
\noindent
All quantitative information about the network geometry follows from structural properties of the adjacency matrix, which can be characterised by a rich variety of different measures~\cite{Albert2002,Newman2003,Boccaletti2006,Costa2007}. We emphasise that although we are considering simple graphs, the resulting graph topology may be highly non-trivial, which justifies the term \textit{complex network} as a contrast to regular chains or lattices.

Recently, it has been suggested that RPs can be alternatively viewed as a complex  network~\cite{Marwan2009,Gao2009,Gao2009b,Donner2010PRE,Donner2010NJP,Donner2010IJBC}, which captures the geometric skeleton of the attractor in pha\-se space (i.e., temporal recurrence relationships cor\-res\-pond to spatial proximity relationships). Specifically, neighbouring observations in phase space are represented by mutually linked vertices of a complex network, i.e.,
\begin{equation}
A_{ij}=A_{ij}(\varepsilon)=R_{ij}(\varepsilon)-\delta_{ij},
\end{equation}
\noindent
where $\delta_{ij}$ is the Kronecker delta used here to avoid self-loops in the network. We note that this setting corresponds to a specific choice of the Theiler window in RQA. As a specific type of networks, the $\varepsilon$-recurrence networks described by the adjacency matrix $A_{ij}$ are geometric graphs~\cite{Penrose2003,Felsner2004} aka spatial networks~\cite{Herrmann2003}, i.e., graphs whose vertices are objects in some metric space (specifically, the phase space of a dynamical system or its reconstruction, e.g., based on time-delay embedding). More specifically, they can be classified as proximity graphs~\cite{Carreira-Perpinan2005} in arbitrary spatial dimensions. As a consequence, given a proper sampling from the considered (dissipative) system, $\varepsilon$-recurrence networks approximate the underlying continuous system and encode relevant geometric information about the corresponding attractor. {Moreover, the quantitative analysis of $\varepsilon$-recurrence networks provides complementary insights with respect to RQA, since corresponding network-theoretic quantities are not based on temporal correlations or explicit time ordering (e.g., line structures in the RP) like most classical RQA measures~\cite{Donner2010NJP}.}

We note that the idea beyond $\varepsilon$-recurrence networks is a straightforward generalisation of recent approaches in neurosciences~\cite{Zhou2006,Zhou2007} and climatology~\cite{Donges2009a,Donges2009b}, where the considered systems are approximated by networks based on mutual correlations between the individual sets of observations measured at certain discrete points in physical space. Furthermore, we emphasise that the\-re are close links to other problems utilising distances between objects in some metric space, such as cluster analysis~\cite{Carreira-Perpinan2005}, dimensionality reduction (e.g., multidimensional scaling~\cite{Borg2005} or isometric feature mapping~\cite{Tenenbaum2000}), or set-oriented approaches for identifying dynamically invariant objects~\cite{Dellnitz2006,Padberg2009}.

It should be mentioned that there are multiple ot\-her approaches to analysing time series by means of complex network methods. Specific concepts proposed so far include transition networks based on a coarse-graining of phase space~\cite{Nicolis2005}, cycle networks~\cite{Zhang2006}, correlation networks~\cite{Yang2008}, visibility graphs~\cite{Lacasa2008}, and $k$-nearest neighbour~\cite{Shimada2008} as well as adaptive nearest-neighbour networks~\cite{Xu2008}. The two last methods differ from the $\varepsilon$-recurrence networks only in the way the local neighbourhood of a vertex is defined. We emphasise that the latter class of methods offers the most general applicability to a variety of different situations (for a detailed discussion of all approaches and their potentials and limits, we refer to~\cite{Donner2010NJP,Donner2010IJBC}). The above mentioned methods have already been used for studying complex systems from various perspectives and fields of applications, however, the evaluation of their full potential is still in the process of exploration. {For the remainder of this paper, we will exclusively consider $\varepsilon$-recurrence networks, implying that the results obtained in this work do not apply to other types of time series networks due to their different construction principles. Specifically, even for $k$-nearest neighbour networks, the construction principle of which is most similar to that of $\varepsilon$-recurrence networks, the local neighbourhood definition is already so different that our considerations cannot be easily transferred to this type of networks. However, since $k$-nearest neighbour networks are based on the original RP definition~\cite{Eckmann1987} and have some further interesting properties, it will be worth considering them in a similar way as done here in future work.}

Recent numerical findings revealed that among ot\-her measures from graph theory, the local and global transitivity properties are particularly well suited for identifying and discriminating qualitatively different types of dynamics. Specifically, the global network transitivity has been demonstrated to provide a good discriminatory statistics for automatically distinguishing between periodic and chaotic dynamics in complex bifurcation scenarios~\cite{Zou2010}. Moreover, it has been sug\-ges\-ted that the local clustering coefficient is able to trace the location of certain dynamically invariant objects \cite{Gao2009,Donner2010NJP}. In this work, we will further elaborate on the relationship between attractor properties on the one hand, and $\varepsilon$-re\-cur\-ren\-ce network properties on the other hand, with a special emphasis on dynamically invariant properties such as fractal dimensions. In particular, we will further explore the interrelationships between local recurrence rates (i.e., the relative frequency of edges a vertex contributes to), the local and global transitivity properties of the graph, and the fractal dimension of the underlying attractor. We note that the latter results in certain scaling properties of different statistical measures that are exclusively based on the temporal evolution of the system under study, establishing a close link between the dynamics on, and the structure of the attractor, which shall be explored here from a complex network perspective. We note that similar findings have also been reported independently concerning links between the structure and function of complex networks~\cite{Boccaletti2006}, e.g., in terms of their synchronisability~\cite{Arenas2008}, the spreading of failures~\cite{Buldyrev2010}, etc.

This paper is organised as follows: In Sec.~\ref{sec2}, we review some basic concepts from complex network as well as dynamical systems theory needed in this work. Some rigorous theoretical results linking the transitivity properties of $\varepsilon$-recurrence networks with the fractal structure of the underlying attractor are provided in Sec.~\ref{sec3} and illustrated by numerical findings in Sec.~\ref{sec4}. Finally, our main results are briefly summarised.

\section{Theoretical background}\label{sec2}

\subsection{Network properties}\label{sec21}

\subsubsection{Direct connectivity properties}\label{sec211}

In complex network research, the most important vertex property is its \textit{degree} (frequently also referred to as degree centrality), i.e., the number of links to other vertices in the graph:
\begin{equation}
k_i=\sum_{j\neq i} A_{ij}.
\end{equation}
\noindent
Since this measure is extensive, i.e., $k_i$ usually increa\-ses monotonously with increasing $N$, we prefer to use a non-extensive property, the normalised degree or local \textit{degree density}, which can be defined as the probability that a given vertex $i$ is linked with any other randomly chosen vertex $j$:
\begin{equation}
\rho_i=P(A_{ij}=1).
\end{equation}
\noindent
For a finite graph, the latter one is estimated by
\begin{equation}
\hat{\rho}_i=\frac{1}{N-1}\sum_{j\neq i} A_{ij}=\frac{k_i}{N-1}.
\end{equation}
\noindent
We note that for $\varepsilon$-recurrence networks, $A_{ij}=A_{ij}(\varepsilon)$ depends explicitly on the recurrence threshold $\varepsilon$ (i.e., the spatial scale of coarse-graining), so that also $k_i$ and $\rho_i$ are functions of $\varepsilon$. In this case, $\hat{\rho}_i(\varepsilon)$ corresponds to the \textit{local recurrence rate} $RR_i(\varepsilon)$ of the observed state $x(t_i)$. In a similar way, we can identify the \textit{global edge density} 
\begin{equation}
\hat{\rho}=\frac{1}{N(N-1)} \sum_{i\neq j} A_{ij}=\frac{1}{N}\sum_{i=1}^N \hat{\rho}_i
\end{equation}
\noindent
of an $\varepsilon$-recurrence network with the (global) \textit{recurrence rate} $RR$ observed for the underlying system. Since in this case, $A_{ij}$ depends explicitly on $\varepsilon$, so do $RR$ as well as all graph-theoretical measures derived from the adjacency matrix.

\subsubsection{Transitivity properties}\label{sec212}

In general, the term transitivity refers to the reproduction of binary relations between mathematical objects. Formally, given a set $X$ of objects, a (directed or undirected) relation $R$ over $X$ is called \textit{transitive} iff whenever $A\in X$ is related to $B\in X$ \textit{and} $B$ is related to $C\in X$, then $A$ is also related to $C$.

{In dynamical systems theory, there are several notions of transitivity, e.g., metric and topological transitivity, which describe properties of certain (continuous) topological transformations. In contrast, in a complex network, the term transitivity is related to fundamental algebraic relationships between triples of discrete objects~\cite{Oxtoby1937,Katok1995}. Specifically, in} graph-theoretical terms, we identify {the set} $X$ with the set of vertices $V$, and {the relation} $R$ with the mutual adjacency of pairs of vertices. Hence, for a given vertex $i\in V$, transitivity refers to the fact that for two other vertices $j,k\in V$ with $A_{ij}=A_{ik}=1$, $A_{jk}=1$ also holds (here as well as in the following general considerations, we omit the $\varepsilon$-dependence of the adjacency matrix for $\varepsilon$-recurrence networks). In a general network, this is typically not the case for all vertices. Consequently, characterising the degree of transitivity (or, alternatively, the relative frequency of closed 3-loops, which are commonly referred to as \textit{triangles}) with respect to some individual vertex or the whole network provides important information on the structural graph properties, which may be related to important general features of the underlying system.

Following the above general considerations, the local transitivity characteristics of a complex network are quantified by the \textit{local clustering coefficient}~\cite{Watts1998}, which measures the probability that two randomly chosen neighbours of a given vertex $i$ are mutually linked,
i.e.,
\begin{eqnarray}
\mathcal{C}_i &=& P(A_{jk}=1|A_{ij}=1,A_{ik}=1) \nonumber \\
&=& \frac{P(A_{ij}=1,A_{ik}=1,A_{jk}=1)}{P(A_{ij}=1,A_{ik}=1)}.
\label{def:cloc}
\end{eqnarray}
For finite graphs, the corresponding probability is typically estimated in terms of the relative frequency of links between the neighbouring vertices of a gi\-ven node $i$, i.e.,
\begin{equation}
\begin{split}
\hat{\mathcal{C}}_i &= \frac{\mbox{number of triangles including vertex } i}{\mbox{number of triples centred on vertex } i} \\
&=\frac{\sum_{j,k} A_{jk} A_{ij} A_{ik}}{k_i(k_i-1)},\label{ciest}
\end{split}
\end{equation}
\noindent
where \textit{triple} refers to a pair $(j,k)$ of vertices that are both linked with $i$, but not necessarily mutually linked.

Local transitivity properties translate to global network properties by sophisticated averaging. In this context, two different measures may be distinguished~\cite{Newman2003,Boccaletti2006,Costa2007}: On the one hand, the \textit{global clustering coefficient} introduced by Watts and Strogatz \cite{Watts1998} is defined as the arithmetic mean of the local clustering coefficients ta\-ken over all vertices of the network:
\begin{equation}
\hat{\mathcal{C}} = \frac{1}{N}\sum_{i=1}^N \mathcal{C}_i = \frac{1}{N}\sum_{i=1}^N \frac{\sum_{j,k=1}^N A_{jk} A_{ij} A_{ik}}{\sum_{j,k=1}^N A_{ij} A_{ik} (1-\delta_{jk})}.
\end{equation}
\noindent
A potential disadvantage of this measure is that it gives equal weights also to vertices with sparse connectivity, which can by definition only contribute to few triangles in the network. On the other hand, the definition of the clustering coefficient according to Barrat and Weigt \cite{Barrat2000,Newman2001}, which has been later suggested to be termed \textit{network transitivity}~\cite{Boccaletti2006}, gives equal weight to all triangles in the network:
\begin{equation}
\begin{split}
\hat{\mathcal{T}}&=\frac{3\times \mbox{number of triangles in the network}}{\mbox{number of linked triples of vertices}} \\
&=\frac{\sum_{i,j,k=1}^N A_{jk} A_{ij} A_{ik}}{\sum_{i,j,k=1}^N A_{ij} A_{ik} (1-\delta_{jk})}. \label{trest}
\end{split}
\end{equation}
\noindent
Note that differences between both measures are ge\-ne\-ric and typically persist even for large networks~\cite{Newman2003}.

\subsubsection{Relationships between different measures}\label{sec213}

As shown above, estimates of both local degree density and local clustering coefficient can be written in terms of (joint) probabilities for the existence of edges in certain parts of a complex network. In terms of $\varepsilon$-recurren\-ce networks (or, even more general, spatially embedded graphs), these different parts may be interpreted as certain regions in (phase) space. Equation~(\ref{ciest}) suggests a possibly nontrivial relationship between the two aforementioned local network properties. For different models of scale-free networks, it has been shown that $\mathcal{C}_i \sim k_i^{-1}$ at least for large degrees $k_i$~\cite{Dorogovstev2002,Szabo2003}. Similar observations have been made for different real-world networks~\cite{Ravasz2002,Ravasz2003,Vazquez2003}. 

\begin{figure*}
\centering\includegraphics[width=0.8\textwidth]{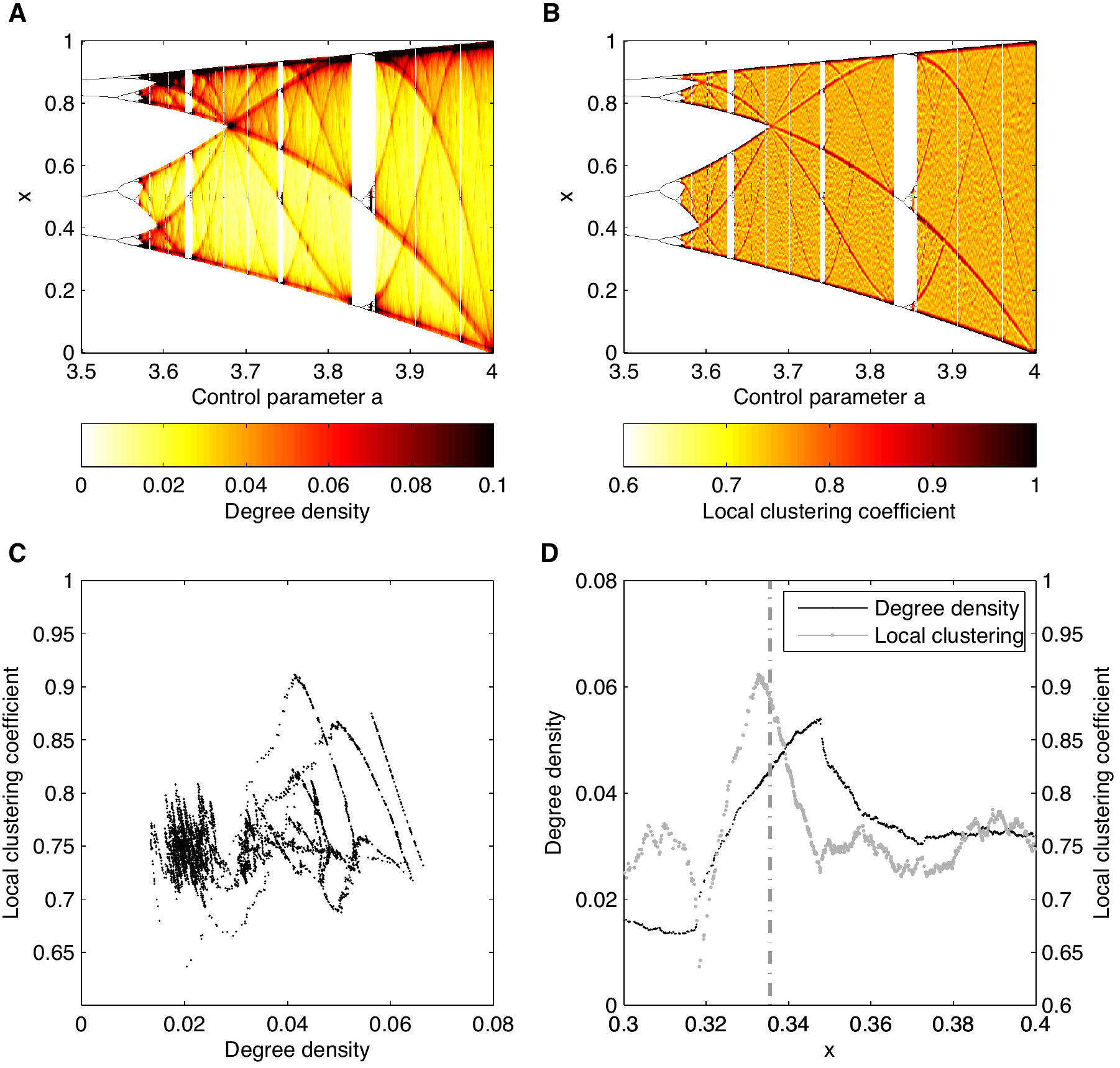}
\caption{Colour-coded representations of estimates of (A) local degree density $\hat{\rho}_i$ and (B) local clustering coefficient $\hat{\mathcal{C}}_i$ for the $\varepsilon$-recurrence networks obtained from trajectories of the logistic map for different control parameters $a$ ($N=10,000$, no embedding, maximum norm, $\varepsilon=0.05\sigma$ with $\sigma$ being the empirical standard deviation of the respectively considered realisations). (C) Scatter plot between $\hat{\rho}_i$ and $\hat{\mathcal{C}}_i$ in $x\in[0.1,0.9]$ (this choice reduces the effects of the attractor boundaries) for $a=3.9$, yielding a rank-order correlation coefficient $\rho_S\approx 0.25$. (D) Magnification of the profile of $\hat{\rho}_i$ (black) and $\hat{\mathcal{C}}_i$ (grey) in the vicinity of a supertrack function (vertical line) for $a=3.9$.}
\label{logmap_clustering}
\end{figure*}

Unlike for the aforementioned examples, for $\varepsilon$-re\-cur\-ren\-ce networks as spatially embedded graphs, the\-re is no obvious simple dependence of $\mathcal{C}_i$ on the density of vertices (and, hence, the local degree density). In contrast, the corresponding correlations between $\mathcal{C}_i$ and $k_i$, which ha\-ve been numerically studied in a previous paper~\cite{Donner2010NJP} for different low-dimensional dynamical systems, have found to be system-specific and often not significant. However, there are examples such as the \textit{logistic map}
\begin{equation}
x_{n+1}=f(x_n)=ax_n(1-x_n)
\label{def:logmap}
\end{equation}
\noindent
with $x\in S\subset[0,1]$ and $a\in[0,4]$, where local maxima of the degree centrality roughly coincide with maxima of the local clustering coefficient (see Fig.~\ref{logmap_clustering}, this feature will be further discussed in Sec.~\ref{sec41} of this paper). In general, the local transitivity properties of $\varepsilon$-recurrence networks do \textit{not} simply capture density effects, but have a distinct meaning in terms of attractor geometry, which will be further highlighted in Sec.~\ref{sec3}.

It should be noted that for $\varepsilon$-recurrence networks, degree centrality, local degree density, and edge density are monotonously increasing functions of $\varepsilon$ by definition. A similar observation has been made for the global clustering coefficient, which follows mainly from the fact that parts of the phase space outside the attractor boundaries get an increasing weight as the proximity threshold $\varepsilon$ becomes larger~\cite{Donner2010NJP}. Note that there is no similar general dependence for the local clustering coefficient. We will also further address this point within the course of this paper.

\subsection{Attractor properties}\label{sec22}

The fundamental structural properties of attractors in dissipative dynamical systems follow from their invariant density $p(x)$ associated with the natural measure $\mu$ as $d\mu(x)=p(x)dx$. The study of the latter is particularly interesting for chaotic systems, where the attractor has a complex shape in phase space, and allows the definition and investigation of dynamically invariant properties such as entropies, fractal dimensions, and related measures of complexity. In this work, we are particularly interested in the concept of fractal dimensions and their relationship with the properties of $\varepsilon$-recurrence networks.

\subsubsection{Global attractor dimensions}\label{sec221}


The most common definition of a fractal dimension ta\-kes a self-similarity property of chaotic attractors into account: Given a partitioning of phase space into a set of fixed $m$-di\-men\-sio\-nal hypercubes of length $l$, one determines the number $n(l)$ of such cubes necessary to fully cover the attractor. Typically, one observes $n(l)\sim l^{D_0}$ with some scaling exponent $D_0$, which is referred to as the \textit{box-counting} or \textit{capacity dimension}
\begin{equation}
D_0=\lim_{l\to 0}\frac{\log n(l)}{\log \frac{1}{l}}.
\end{equation}

Although $D_0$ is often called ``the'' fractal dimension, there is a multiplicity of other definitions of fractal dimensions. These \textit{generalised dimensions} are related with the order-$q$ R\'enyi entropies
%
\begin{equation}
S_q(l)=\frac{1}{1-q}\log\sum_{i=1}^{n(l)} p_l(i)^q \quad (q\in\mathbb{R}^+\cup\{0\}),
\end{equation}
\noindent
as~\cite{Grassberger1983PLA}
\begin{equation}
D_q=\lim_{l\to 0} \frac{S_q(l)}{\log\frac{1}{l}}
\label{def:dq}
\end{equation}
\noindent
with $D_{q_1}\leq D_{q_2}$ for $q_1>q_2$. Special cases of this definition include $D_0$, the information dimension $D_1$ (where $S_1$ corresponds to the Shannon entropy), and the correlation dimension $D_2$. The latter can be equivalently defined using the 
%
%
%
\textit{correlation integral}~\cite{Grassberger1983PRL}
\begin{equation}
C(\varepsilon)=\int d\mu(x)\int d\mu(y)\ \Theta(\varepsilon-\|x-y\|), \label{corrint}
\end{equation}
\noindent
as
\begin{equation}
D_2=\lim_{\varepsilon\to 0} \frac{\log C(\varepsilon)}{\log\epsilon}.
\label{defd2}
\end{equation}
Given only a finite set of sampled points on a specific trajectory (i.e., a time series) of the system, an unbiased estimator of $C(\varepsilon)$ is given by the \textit{correlation sum}
\begin{equation}
\hat{C}(\varepsilon)=\lim_{N\to\infty} \frac{1}{N(N-1)}\sum_{i,j=1{,i\neq j}}^N \Theta(\varepsilon-\|x_i-x_j\|), \label{corrsum}
\end{equation}
\noindent
which corresponds to the global recurrence rate $RR$ (and, hence, the edge density $\hat{\rho}$ of the associated $\varepsilon$-recurrence network). If the supposed power-law be\-ha\-viour $C(\varepsilon)\propto\varepsilon^{D_2}$
%
of the correlation integral applies, one may usually find that also $\hat{\rho}(\varepsilon)\propto \varepsilon^{D_2}$ for $\varepsilon\in [\varepsilon_{min},\varepsilon_{max}]$, such that $D_2$ can be estimated as 
\begin{equation}
\hat{D}_2=\frac{d\log\hat{\rho}(\varepsilon)}{d\log\varepsilon} \quad \mbox{for} \quad \varepsilon\in [\varepsilon_{min},\varepsilon_{max}]. \label{estd2}
\end{equation}
\noindent
The finite size of this scaling interval is caused by the fact that (i) for small $\varepsilon$, the number of recurrences (i.e., neighbours in phase space with respect to the considered $\varepsilon$-distance) is too low to observe the correct slope of $\log\hat{\rho}=f(\log\varepsilon)$ with sufficient statistical confidence (fi\-ni\-te-resolution limit of a finite time series), whereas (ii) for large $\varepsilon$, parts of phase space that are not covered by the attractor (and therefore do not contain any observations) are successively included in the considered $\varepsilon$-balls, which leads to a saturation of $\hat{\rho}$. Moreover, in the latter case, more and more redundant information is included in the estimation (e.g., due to the considerations of points that are nieghbours in phase space just because they are also close in time). 
Therefore, in most available methods for estimating $\hat{D}_2$ from time series, the consideration of a sufficiently large ensemble of different values of $\varepsilon$ is necessary to obtain proper estimates.

A practical alternative is considering scale-local dimensions $\hat{D}_q(\varepsilon)$~\cite{Reid2003}, i.e., numerical estimates of $\hat{D}_q$ obtained from the respective relationships evaluated for only one \textit{fixed} value of $\varepsilon$. Note, however, that these quantities often do not provide proper estimates of the \textit{true} dimension. 

\subsubsection{Pointwise dimensions}\label{sec223}

The defining equation~(\ref{defd2}) of the correlation dimension has been formulated globally as a property of the whole attractor. However, the correlation sum is defined as an unweighted average of contributions from all observed state vectors $x_i$, i.e.,
\begin{equation}
\hat{C}(\varepsilon)=\frac{1}{N}\sum_{i=1}^N \hat{C}_i(\varepsilon)
\end{equation}
\noindent
with
\begin{equation}
\hat{C}_i(\varepsilon)=\frac{1}{N-1}\sum_{j=1,{j\neq i}}^N \Theta(\varepsilon-\|x_i-x_j\|)=\hat{\rho}_i(\varepsilon)
\end{equation}
\noindent
coinciding with the local recurrence rate (local degree density $\hat{\rho}_i$) of $x_i$ (the same considerations apply to the correlation integral $C(\varepsilon)$ in Eq.~(\ref{corrint})). Motivated by this, one defines the
%
pointwise dimension~\cite{Farmer1983}
\begin{equation}
D_p(x)=\lim_{\varepsilon\to 0} \frac{\log\mu(B_\varepsilon(x))}{\log\varepsilon},
\end{equation}
\noindent
{where $\mu(B_\varepsilon(x))$ is the measure of a ball of radius $\varepsilon$ centred at $x$. Commonly, this pointwise dimension is estimated as}
\begin{equation}
\hat{D}_p(x_i)=\frac{d\log\hat{\rho}_i(\varepsilon)}{d\log\varepsilon} \quad \mbox{for} \quad \varepsilon\in [\varepsilon_{min},\varepsilon_{max}].
\label{eq_dpest}
\end{equation}
\noindent
If $D_p(x)$ exists (note that convergence cannot be expected in general), it is independent of $x$ for almost all $x$ with respect to the invariant measure $\mu$. Since unlike for $C(\varepsilon)$, no two-point correlations are involved in the local recurrence rate, the pointwise dimension provides a local estimate of the information dimension $D_1$ rather than $D_2$. As for the (global) correlation dimension, the practical estimation of pointwise dimensions may be challenging since the limit $\varepsilon\to 0$ can hardly be assessed with a limited number of data available. 



\subsubsection{Lyapunov dimension}\label{sec224}

The Lyapunov (or Kaplan-Yorke) dimension is based on the temporal stretching and folding characteristics of the dynamical system under study. Let $\{\lambda_1, \dots, \lambda_m\}$ denote the spectrum of Lyapunov exponents of the system ordered such that $\lambda_1\geq \lambda_2 \geq \cdots \geq \lambda_m$, and $n$ be the largest integer such that
\begin{equation}
\sum_{j=1}^n \lambda_j \geq 0 
\label{eq:lyapunov_dim_condition}. 
\end{equation}
Then the \textit{Lyapunov dimension} or \textit{Kaplan-Yorke dimension} $D_L$ is defined as~\cite{Farmer1983,Kaplan1979,Ott1993}
\begin{equation}
D_L = n + \frac{1}{|\lambda_{n+1}|} \sum_{j=1}^n \lambda_j. 
\label{eq:lyapunov_dim_def}
\end{equation}
The Kaplan-Yorke conjecture states that $D_1=D_L$ for typical attractors. 

The Lyapunov dimension is typically considered a global measure as it applies to the whole attractor except for a set of measure zero. In contrast, the \textit{local Lyapunov dimension} $D_L(x)$ can vary across the attractor~\cite{Hunt1996,Gelfert2003}. It has (so far) mainly been studied for maps $F$ and can be obtained from Eqs. (\ref{eq:lyapunov_dim_condition}, \ref{eq:lyapunov_dim_def}) after the replacements $D_L \to D_L(F,x)$ and $\lambda_j \to \ln\left(a_j(x)\right)$, where $a_j(x)$ are the ordered singular values of the map's Jacobian $DF(x)$. The most important property of $D_L(F, x)$ is that it provides an upper bound on the box-counting dimension $D_0$, i.e.,
\begin{equation*}
D_0 \leq \max_x D_L(F,x).
\end{equation*}
The local Lyapunov dimension $D_L(F^k,x)$ of the $k$-th iterate of the map $F$ converges to the Lyapunov dimension $D_L$ as $k\to \infty$~\cite{Hunt1996}.

\subsection{Links between network and attractor properties}\label{sec23}

We have already argued that (given a proper sampling of the considered trajectory) the local degree density of an $\varepsilon$-recurrence network (i.e., the local recurrence rate) allows defining an estimator of the invariant density $p(x)$ of the underlying attractor. The latter is known to be fundamental for both the structure of, and the dynamics on the attractor, whereas a similar statement applies to the connectivity of vertices and the overall structure of a network. These established links suggest that there might well be more possible interrelationships between dynamical system and network properties. This is evident for the estimation of the correlation dimension from the scaling of the edge density with varying $\varepsilon$. In addition, as it will be shown in a complementary paper~\cite{Zou2011}, there are cases where the degree distribution of an $\varepsilon$-recurrence network includes a scale-free part, the characteristic exponent of which can be related to the pointwise dimension of the underlying system. We note that similar observations have been made for another type of complex networks generated from time series, the so-called visibility graphs, which also show a power-law decay of the degree distribution for certain fractal time series, the exponent of which is directly related to the Hurst parameter~\cite{Lacasa2009,Ni2009}.

Another important relationship between the geometric properties of a dynamical system and the transitivity properties of their induced $\varepsilon$-recurrence networks can be inferred from the theory of random geometric graphs~\cite{Dall2002}. Specifically, it has been shown that for such graphs, the expected global clustering coefficient depends on the (integer) dimension of the metric space in which the considered graph is embedded. In the remainder of this paper, we will generalise this result to arbitrary non-integer spatial dimensions and discuss the resulting implications in some detail.

\section{Continuous clustering coefficient and clustering dimension}\label{sec3}

In the following, we develop and successively apply a general theory linking the local as well as global transitivity properties of $\varepsilon$-recurrence networks with geometric attractor properties. For this purpose, we define a novel notion of fractal dimension based on these transitivity properties, and compare these new definitions with several other existing measures descri\-bed in the previous section.

\subsection{Continuous measures for transitivity}\label{sec31}

\subsubsection{General theory}\label{sec311}

Let $p$ be a probability density on some set $S\subseteq R^m$, such that all closed $\varepsilon$-balls 
$$B_\varepsilon(x)=\{y\in S:d(x,y)\leq\varepsilon\}$$ 
with $x\in\bar S$ and $\varepsilon>0$ are $p$-measurable, where 
$$d(x,y)=\max_{i=1,\dots,m} |x_i-y_i|$$ 
is the maximum metric and $\bar S$ is the topological closure of $S$. Then we define the {\em continuous $\varepsilon$-degree density} and {\em continuous $\varepsilon$-clustering coefficient} of any point $x\in\bar S$ as
\begin{eqnarray}
	\rho(x;\varepsilon) &=& \int_{B_\varepsilon(x)}p(y)\,dy = \int_{B_\varepsilon(x)} d\mu(x) \label{def:contrho} \\
	\mathcal{C}(x;\varepsilon) &=& \frac{\int\!\!\!\int_{B_\varepsilon(x)} p(y) p(z) \Theta(\varepsilon-d(y,z))\,dy\,dz}{\rho(x;\varepsilon)^2}, \label{def:contcloc}
\end{eqnarray}
the latter being the probability that two points $y$ and $z$ randomly drawn according to $p$ are closer than $\varepsilon$ given they are both closer than $\varepsilon$ to $x$. As a global measure, we define the {\em continuous $\varepsilon$-transitivity} of $S$ as 
\begin{eqnarray}
\mathcal{T}(\varepsilon) &=& \bigg[ \int\!\!\!\int\!\!\!\int_S p(x) p(y) p(z) \Theta(\varepsilon-d(x,y)) \times\nonumber \\
&& \qquad \times \Theta(\varepsilon-d(y,z)) \Theta(\varepsilon-d(z,x))\,dx\,dy\,dz \bigg] \bigg/ \nonumber \\
&& \quad \bigg[ \int\!\!\!\int\!\!\!\int_S p(x) p(y) p(z) \Theta(\varepsilon-d(x,y)) \times\nonumber \\ && \qquad \times \Theta(\varepsilon-d(z,x))\,dx\,dy\,dz \bigg],
\end{eqnarray}
which is the probability that among three points $x,y,z$ drawn randomly according to $p$, $y$ and $z$ are closer than $\varepsilon$ given they are both closer than $\varepsilon$ to $x$.

\subsubsection{Examples: Dynamical systems defined on the unit interval}\label{sec312}

If $S$ is the unit box $[0,1]^m$ and $p$ is the uniform density $p(x)\equiv 1$, one can easily see that $\mathcal{C}(x;\varepsilon)=\mathcal{T}(\varepsilon)=(3/4)^m$ for all points $x\in[\varepsilon,1-\varepsilon]^m$. This is because for $m=1$, a randomly chosen $y\in B_\varepsilon(x)=[x-\varepsilon,x+\varepsilon]$ has on average three quarters of $B_\varepsilon(x)$ within its own $B_\varepsilon(y)$, namely half of $B_\varepsilon(x)$ for $y=x\pm\varepsilon$ and the whole $B_\varepsilon(x)$ for $y=x$. For $m>1$, the integral in the numerator of Eq.~(\ref{def:contcloc}) decomposes as 
\begin{equation}
\begin{split}
\mathcal{C}(x;\varepsilon)& = \frac{1}{(2\varepsilon)^{2m}} \prod_{i=1}^m \int\!\!\!\int_{x_i-\varepsilon}^{x_i+\varepsilon}\Theta(\varepsilon-|\eta-\zeta|)\,d\eta\,d\zeta \\
& =(3\varepsilon^2)^m/(2\varepsilon)^{2m}=(3/4)^m.
\end{split}
\end{equation}
\noindent

The same is true whenever $x$ is in the topological interior of $\bar S$, $p$ is sufficiently smooth in a neighbourhood of $x$, and $\varepsilon$ is small, because then the situation looks locally approximately the same as for the uniform density on the unit box. For example, if $S$ and $p$ are the main attractor and invariant density of the logistic map (\ref{def:logmap}) at $a=4$, for which $S=(0,1)$ and~\cite{Sprott2003}
\begin{equation}
p(x)=\frac{1}{\pi\sqrt{x(1-x)}},
\label{logmap_density}
\end{equation}
\noindent
then one can also show that for small $\varepsilon$, $$\rho(x;\varepsilon)\approx \frac{2\varepsilon}{\pi\sqrt{x}}$$ and 
$$\int\!\!\!\int_{B_\varepsilon(x)} p(y) p(z) \Theta(\varepsilon-d(y,z))\,dy\,dz\approx \frac{3\varepsilon^2}{\pi^2 x},$$ so that $$\mathcal{C}(x;\varepsilon)\to 3/4$$ as $\varepsilon\to 0$. For $x\in\{0,1\}$, we get $\mathcal{C}(x;\varepsilon)=1$ for all $\varepsilon$, since then all pairs of points in $B_\varepsilon(x)$ are on the same side of $x$ and thus have a distance $\leq\varepsilon$. Note that these results are consistent with recent findings described in \cite{Donner2010NJP} (see Sec.~\ref{sec41}).

In general, the stronger a smooth $p$ varies inside $B_\epsilon(x)$, the more $\mathcal{C}(x;\epsilon)$ exceeds its lower bound $(3/4)^m$. More precisely, the larger $m$, the closer $\mathcal{C}(x;\epsilon)$ is related to the 2nd-order R\'enyi entropy of $p$ restricted to $B_\epsilon(x)$, since the Heaviside function then acts more and more like a delta function and the denominator of $\mathcal{C}(x;\epsilon)$ becomes approximately proportional to $\int_{B_\epsilon(x)} p(y)^2\ dy$. The phenomenon that the clustering coefficient is influenced by the local density {\em variability} rather than by the local density {\em level} was also observed in a different context for climate networks~\cite{Heitzig2010}. 

We have to emphasise that the above result $\mathcal{C}(x;\varepsilon)=\mathcal{T}(\varepsilon)=(3/4)^m$ is only valid when using the maximum norm for defining distances between points in phase space. For other choices of the metric $d$, e.\,g., the Euclidean one, there will in general not be a similarly simple exponential relationship, although it is sometimes at least asymptotically exponential. For example, for the uniform density on the unit box and using the Euclidean metric, we find
\begin{equation}
\begin{split}
\mathcal{C}(x;\varepsilon) = & \ 1 - \frac{m\Gamma(m/2)}{2\sqrt\pi\Gamma((m+1)/2)}
\bigg[{}_2F_1\left(\frac 1 2,\frac{1-m}2;\frac 3 2; \frac 1 4\right) \\
& \qquad - \frac 1{m+1} {}_2F_1\left(\frac{1-m}2,\frac{m+1}2;
				\frac{m+3}2;\frac 1 4\right) \bigg] \\
\approx & \ 3\sqrt{\frac{2}{\pi m}} \left(\frac{3}{4}\right)^{(m+1)/2} \approx 0.862^m/3,
\end{split}
\end{equation}
\noindent
where ${}_2F_1(\cdot)$ is the hypergeometric function and the exponential fit is only good for $m>40$. The latter result can be obtained from the expressions for the volumes of $m$-dimensional spheres and their intersections and is consistent with previous findings in~\cite{Dall2002}.

\subsection{Clustering dimensions}\label{sec32}

\subsubsection{General theory}\label{sec321}

The aforementioned observations motivate the definition of two new local and two new global measures of dimension for general $S$ and $p$, namely the {\em upper} and {\em lower clustering dimension} of $S$ at $x$,
\begin{eqnarray}
	D_\mathcal{C}^u(x) &=& \limsup_{\varepsilon\to 0}\frac{\log \mathcal{C}(x;\varepsilon)}{\log(3/4)}\\
	\mbox{and}~
	D_\mathcal{C}^l(x) &=& \liminf_{\varepsilon\to 0}\frac{\log \mathcal{C}(x;\varepsilon)}{\log(3/4)},
\end{eqnarray}
and the {\em upper} and {\em lower transitivity dimension} of $S$,
\begin{eqnarray}
	D_\mathcal{T}^u &=& \limsup_{\varepsilon\to 0}\frac{\log \mathcal{T}(\varepsilon)}{\log(3/4)}\\
	\mbox{and}~
	D_\mathcal{T}^l &=& \liminf_{\varepsilon\to 0}\frac{\log \mathcal{T}(\varepsilon)}{\log(3/4)}.
\end{eqnarray}
\noindent

Note that $D_\mathcal{C}^u(x)$ and $D_\mathcal{C}^l(x)$ need not be continuous in $x$. For example, in the logistic map with $a=4$, the border points $x\in\{0,1\}$ have $\mathcal{C}(x;\varepsilon)=1$ (see above), so that $D_\mathcal{C}^u(x)=D_\mathcal{C}^l(x)$ jumps from $1$ to $0$ for $x\to 0,1$. For $a<4$, such discontinuities also occur in the interior of $\bar S$, because inside $B_\varepsilon(x)$ for a point $x$ on a supertrack function~\cite{Oblow1988,marwan2002herz} of the logistic map, $p$ has asymptotically a power-law form $$p(y)\sim |y-x|^\gamma$$ on one side of $x$, and is approximately constant on the other side of $x$~\cite{Ott1993}. To see this, assume that the $k$-th iterate of the map, $f^k$, has a local maximum at $x$, $f^k(x)=y$. Then, all points close to $x$ get mapped by $f^k$ to points rather close to $y$ but \textit{smaller} than $y$. This explains why the density then has a peak at $y$, which diverges only to the left side of $y$. The power-law decay can be seen from the local \textit{quadratic} approximation of $f^k$ at $x$, since that has a slope linear in the distance from $x$, and the density transforms using the inverse of the slope, i.e., 1/(distance from $x$). As a consequence, for $\varepsilon\to 0$, almost all mass inside $B_\varepsilon(x)$ is on the power-law side and, hence, almost all $y,z\in B_\varepsilon(x)$ have distance $\leq\varepsilon$, yielding $\mathcal{C}(x;\varepsilon)\to 1$ for $\varepsilon\to 0$. Hence, $D_\mathcal{C}^u(x)=D_\mathcal{C}^l(x)=0$ at all points on the supertrack functions of the logistic map. This behaviour can be seen in Fig.~\ref{logmap_clustering}, where the supertrack functions are clearly visible in terms of pronounced maxima of $\mathcal{C}(x)$ that have been numerically estimated from the $\varepsilon$-recurrence networks of sample trajectories~\cite{Donner2010Nolta}.

\subsubsection{Examples: Self-similar sets}\label{sec322}

If $S$ is highly self-similar, $D_\mathcal{C}^u(x)$ and $D_\mathcal{C}^l(x)$ can also differ considerably, as can be seen from the example in which $p(x)\equiv 1$ and $S=\{x\in[-1,1]:|x|\in S_C\}$, where $S_C\subset[0,1]$ is the \textit{Cantor set} (= numbers that have a ternary expansion in which all digits are either 0 or 2). In that case, it is easy to see from the symmetry that $\mathcal{C}(0;\varepsilon)$ oscillates between $3/4$ and $1$ for $\varepsilon\to 0$, with $\mathcal{C}(0;\varepsilon)=3/4$ if $\varepsilon=1/3^n$ for some integer $n\geq 0$, and $\mathcal{C}(0;\varepsilon)=1$ if $\varepsilon=2/3^n$ for some integer $n\geq 1$. Hence, $D_\mathcal{C}^u(0)=1$ and $D_\mathcal{C}^l(0)=0$. Similarly, for almost all $x\in S$, one can show that $\mathcal{C}(x;\varepsilon)$ is infinitely often $7/8$ and $1$ for $\varepsilon\to 0$, so that $D_\mathcal{C}^u(x)\geq\log(7/8)/\log(3/4)\approx 0.464$ and $D_\mathcal{C}^l(x)=0$. Only for the countably many points of the form $x=\pm a/3^n$ with integers $a,n>0$, we get $\mathcal{C}_\varepsilon(x)=1$ for all $\varepsilon<1/3^n$ and, hence, $D_\mathcal{C}^u(x)=D_\mathcal{C}^l(x)=0$. In the same example, also $D_\mathcal{T}^u$ and $D_\mathcal{T}^l$ differ since $\mathcal{T}(\varepsilon)=1$ whenever $\varepsilon=1/3^n$ with integer $n>0$ and $\mathcal{T}(\varepsilon)=11/13$ whenever $\varepsilon=5/3^n$ with integer $n>1$, with intermediate values for other values of $\varepsilon$, so that $D_\mathcal{T}^u=\log(11/13)/\log(3/4)\approx 0.581$ and $D_\mathcal{T}^l=0$.

The same values are obtained when $S$ is a version of the Cantor set in which, starting with the unit interval, iteratively the middle fraction of relative width $1-2\alpha$ is removed from every remaining interval, for $\alpha\leq 1/3$, where $\alpha=1/3$ gives the standard Cantor set. Other measures of dimension, however, have values depending on $\alpha$, e.\,g., the box-counting, information, and correlation dimension of $S$ and the pointwise dimension of almost all $x\in S$ are all $-\log 2/\log\alpha$, which varies between $\log 2/\log 3\approx 0.631$ and $0$. In particular, this shows that both $D_\mathcal{C}^u(x)$ and $D_\mathcal{T}^u$ can be either smal\-ler or larger than all those classical measures of dimension. This finding is supported by further analytical as well as numerical results (see, e.g., Sec.~\ref{sec422}).

As a further example, let us consider the \textit{generalised baker's map}
\begin{equation}
\begin{split}
x_{n+1} &= \left\{ \begin{array}{ll} \lambda_a x_n, & \quad y_n<\alpha,\\ (1-\lambda_b) + \lambda_b x_n, & \quad y_n>\alpha, \end{array} \right. \\
y_{n+1} &= \left\{ \begin{array}{ll} y_n/\alpha, & \quad y_n<\alpha,\\ (y_n-\alpha)/(1-\alpha), & \quad y_n>\alpha \end{array} \right.
\end{split}
\label{genbakersmap}
\end{equation}
\noindent
with $\alpha<1$, $\lambda_a,\lambda_b>0$ and $\lambda_a+\lambda_b\leq 1$~\cite{Farmer1983,Ott1993}, which yields a transformation of the unit square $[0,1]\times[0,1]$. If $S$ is the attractor of the symmetric version of this map with $\alpha=1/2$ and $\lambda_a=\lambda_b=1/4$), it is a cartesian product of a Cantor set similar to those discussed above and the unit interval. Hence, for almost all $x\in S$, $\mathcal{C}(x;\varepsilon)$ attains $7/8\cdot 3/4$ and $1\cdot 3/4$ infinitely often for $\varepsilon\to 0$, so that $D_\mathcal{C}^u(x)\geq \log(7/8)/\log(3/4)+1\approx 1.464$ and $D_\mathcal{C}^l(x)=1$. Similarly, we obtain that $D_\mathcal{T}^u=\log(11/13)/\log(3/4)+1\approx 1.581$ and $D_\mathcal{T}^l=1$.

The above examples suggest that it might be interesting to take the difference $D_\mathcal{T}^u-D_\mathcal{T}^l$ as a measure of self-similarity of an attractor. As a counter-example, one may consider again the chaotic attractor of the logistic map at $a=4$, for which $D_\mathcal{C}^{u,l}(x)=D_\mathcal{T}^{u,l}\equiv D_\mathcal{T}=1$ is consistent with $D_p(x)=D_1=m=1$ as one would expect for this non-fractal case.

\subsubsection{Clustering and topological dimension of phase space}\label{sec323}

In some cases, $D_\mathcal{C}^u(x)$ and $D_\mathcal{C}^l(x)$ may even exceed the non-fractal dimension $m$ of the surrounding space. For example, for $S=[-1,1]$ and $p(x)=3x^2/2$, $D_\mathcal{C}^l(x)=D_\mathcal{C}^u(x)=m=1$ for all $x\neq 0$, but $D_\mathcal{C}^l(0)=D_\mathcal{C}^u(0)\approx 2.24>m$ since $\mathcal{C}(0;\varepsilon)=21/40<3/4$ for all $\varepsilon\in(0,1)$. This is because of $p(0)=0$, so $p(y)$ cannot be considered approximately constant on $B_\varepsilon(0)$ for small $\varepsilon$, although $p$ is smooth. The best upper bound for $D_\mathcal{C}^u(x)$ in terms of $m$ is $m\log 2/(\log 4-\log 3)\approx 2.41\,m$. To see this, we split $B_\varepsilon(x)$ into $2^m$ generalised ``quadrants'' corresponding to the possible vectors of signs of $y_i-x_i$, $i=1\dots m$. Note that all pairs $y,z\in B_\varepsilon(x)$ that are in the same generalised quadrant have $d(y,z)\leq\varepsilon$, and the probability of $y,z$ being in the same generalised quadrant is at least $1/2^m$ no matter how $p(B_\varepsilon(x))$ is distributed over these regions, so that $\mathcal{C}(x;\varepsilon)\geq 1/2^m$. That this bound is sharp even for arbitrarily smooth $p$ can be seen from the set of examples with $S=[-1,1]^m$ and $p(x)=\prod_{i=1}^m(2a+1)x_i^{2a}/2$ with integers $a,m>0$, for which $\mathcal{C}(0;\varepsilon)\to 1/2^m$ as $a\to\infty$.

\begin{figure}
\begin{center}
\includegraphics[width=0.3\textwidth]{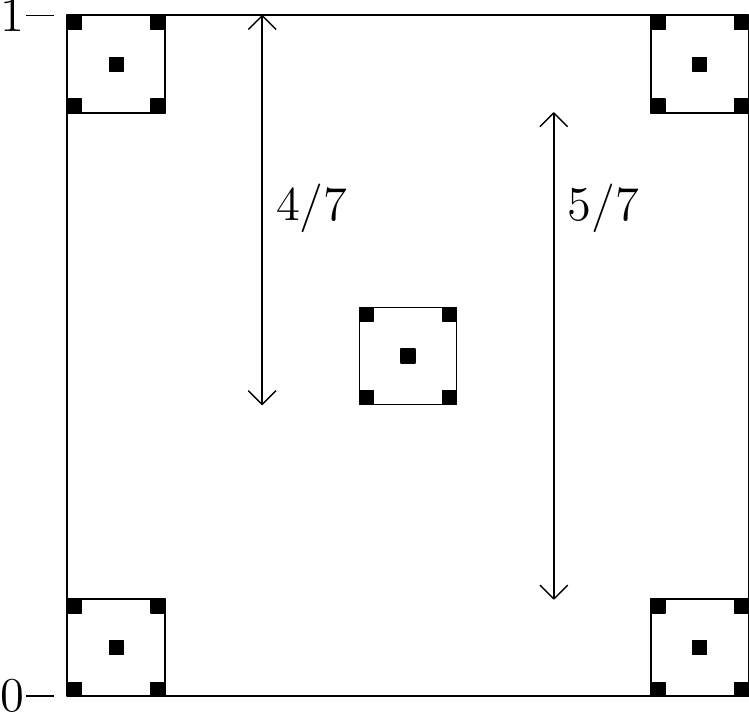}
\end{center}
\caption{\label{fig:cubes} Intermediate step $S_2$ (black squares) of the fractal construction to show that $D_\mathcal{T}^u$ can exceed $m$.}
\end{figure}

That $D_\mathcal{T}^u$ might also exceed $m$ can be seen from another Cantor-like example. Starting with the unit cube $S_0=[0,1]^m$, replace this cube by $1+2^m$ smaller cubes of the form $\prod_{i=1}^m [3/7+s_i,4/7+s_i]$ with $s_1=\cdots=s_m=0$ or $s_i\in\{-3/7,3/7\}$ for all $i$, i.\,e., one small cube located at the center of the original cube, and $2^m$ cubes fit into the corners of the original cube, giving a set $S_1\subset S_0$. Repeating the same replacement infinitely often with each cube gives a descending sequence of sets $S_k$. Figure~\ref{fig:cubes} shows the result $S_2$ of two iterations for $m=2$. The intersection $S=\bigcap_{i=1}^\infty S_i=S_\infty$ is a Cantor-like fractal for which $\mathcal{T}(\varepsilon)=(7\cdot 2^m+1)/(4^m+6\cdot 2^m+1)<(3/4)^m$ whenever $m\geq 5$ and $\varepsilon = 4/7^n$ for some integer $n>0$, hence $D_\mathcal{T}^u(S)>m$. We do not know whether even $D_\mathcal{T}^l(S)$ can exceed $m$ but conjecture that this is impossible (this conjecture is supported by further numerical results in Sec.~\ref{sec4}).

\subsubsection{Local clustering and Lyapunov dimensions}\label{sec324}

For a homogeneous fractal the pointwise dimension as the most traditional other local dimension measure by definition equals the global measure of information dimension for all $x$ except for a set of measure zero, i.e., $D_p(x)=D_p=D_1$~\cite{Farmer1983}. However, many chaotic attractors have a complex internal structure (e.g., embedded objects of measure zero with deviating pointwise dimensions), such as the R\"ossler~\cite{Veronig2000} and Lorenz systems~\cite{Veronig2000,Gratrix2004}. In such cases, $D_\mathcal{C}^u(x)$ and $D_\mathcal{C}^l(x)$ can show considerable regional differences (see Sec.~\ref{sec43} for the R\"ossler system). Beside the pointwise dimensions, the only other local measure of dimension with this property that we are aware of is the local Lyapunov dimension $D_L(x)$~\cite{Hunt1996}, which is based on the local contraction of a map. Thus, it might be worthwile to compare these measures. 

For the logistic map, e.\,g., $D_L(x)=1$ for $x\in[3/8,5/8]$ and $D_L(x)=0$ otherwise, while $D_\mathcal{C}^u(x)=D_\mathcal{C}^l(x)=1$ for $x\in (0,1)$ and $D_\mathcal{C}^u(x)=D_\mathcal{C}^l(x)=0$ for $x\in\{0,1\}$. For the generalised baker's map, we will provide some results in Sec.~\ref{sec422}.

\subsection{Estimation of clustering dimensions}\label{sec33}

In general, if $S$ is an attractor of an ergodic dynamical system with invariant density $p$,
we can estimate $\mathcal{C}(x(t_i);\varepsilon)$ and $\mathcal{T}(\varepsilon)$ from a finite sample $x(t_i)\in S$ of a trajectory in $S$ sampled at time points $t_1,\dots,t_N$, using the standard (sample) clustering coefficients (Eq.~(\ref{ciest})) and transitivity (Eq.~(\ref{trest})).


By the weak law of large numbers, $\hat{\mathcal{C}}_i(\varepsilon)$ and $\hat{\mathcal{T}}({\varepsilon})$ converge in probability to $\mathcal{C}(x(t_i);\varepsilon)$ and $\mathcal{T}(\varepsilon)$ as $N\to\infty$, for each $\varepsilon>0$ and a general choice of time points $t_i$ (e.\,g., regularly spaced with a time step that is coprime with all periodic orbits' periods). 
In other words, $\hat{\mathcal{C}}_i(\varepsilon)$ and $\hat{\mathcal{T}}({\varepsilon})$ are statistically {\em consistent} estimators of $\mathcal{C}(x(t_i);\varepsilon)$ and $\mathcal{T}(\varepsilon)$. Our four new dimension measures can then be estimated from a sufficiently long sampled trajectory as
\begin{eqnarray}
	\hat{D}_{\mathcal{C},i}^u &=& \max_{\varepsilon\in \mathcal{E}}\frac{\log \hat{\mathcal{C}}_i(\varepsilon)}{\log(3/4)},\\
	\hat{D}_{\mathcal{C},i}^l &=& \min_{\varepsilon\in \mathcal{E}}\frac{\log \hat{\mathcal{C}}_i(\varepsilon)}{\log(3/4)},\\
	\hat{D}_{\mathcal{T}}^u &=& \max_{\varepsilon\in \mathcal{E}}\frac{\log \hat{\mathcal{T}}(\varepsilon)}{\log(3/4)},\\
	\mbox{and}~
	\hat{D}_{\mathcal{T}}^l &=& \min_{\varepsilon\in \mathcal{E}}\frac{\log \hat{\mathcal{T}}(\varepsilon)}{\log(3/4)}
\end{eqnarray} 
for a suitably large set $\mathcal{E}$ of different values of $\varepsilon$ that are as small as possible while still providing for sufficiently large values of $k_i(\varepsilon)$. We emphasise that the above set of equations is only feasible if $\hat{\mathcal{C}}_i(\varepsilon)>0$ and $\hat{\mathcal{T}}(\varepsilon)>0$, respectively. For all other points in phase space represented by a vertex $i$, the aforementioned dimension measures cannot be defined in a meaningful way.

Note that $k_i(\varepsilon)$ has a binomial distribution and becomes (for $\varepsilon\to 0$ and large $N$) asymptotically Poissonian with mean and variance~\cite{Herrmann2003} $$\lambda\sim Np(x)\varepsilon^{D_p(x)},$$ where $D_p$ again denotes the standard pointwise dimension. For a given $k_i(\varepsilon)$, $C_i(\varepsilon)$ has asymptotically the mean $C_{\varepsilon}(x(t_i))$ and a variance proportional to $k_i(\varepsilon)^{-2}$, so that also $C_i(\varepsilon)$ has asymptotically a variance proportional to $k_i(\varepsilon)^{-2}$. For this reason vertices with low degree are unlikely to yield reliable estimates of local clustering dimensions (see below).


The quotient $\max \mathcal{E}/\min \mathcal{E}$ should exceed the magnification factor of any suspected self-similarity (i.\,e., 3 in the Cantor set), see Fig.~\ref{fig:gbm_cdim}B for an example. {In general, it has been established that there is no generally applicable rule for choosing $\varepsilon$ for computing recurrence network properties~\cite{Donner2010PRE}. In fact, the values of $\varepsilon$ that may provide feasible results are restricted by practical considerations, i.e., the available sample size determines the smallest possible spatial dimensions of structures to be resolved by network-theoretic measures, which directly relates to the minimally possible $\varepsilon$, whereas large $\varepsilon$ do not allow obtaining information on the geometric fine structures of the attractor. In typical situations, a reasonable trade-off can be found for edge densities below 5\%. With respect to the joint estimation of upper and lower clustering/transitivity dimensions, our numerical studies (see Sec.~\ref{sec4}) reveal that for attractors with a self-similar structure, these measures typically alternate between lower and upper bounds as $\varepsilon$ is varied (cf.~Fig.~\ref{fig:gbm_cdim}B). According to this observation, we suggest as a rule of thumb that the range of $\varepsilon$ should be chosen so that both upper and lower limits can be identified from at least two distinct intervals of $\varepsilon$, respectively.} Note that in contrast to the estimation of other notions of dimension (like with the Grassberger-Procaccia algorithm in the case of correlation dimension \cite{grassberger83}), it is not necessary to estimate the slope from a double-lo\-ga\-rith\-mic plot here.

\section{Examples}\label{sec4}

In the following, we illustrate our previous considerations by numerical as well as further analytical results obtained for some benchmark examples of both low-dimensional maps and time-continuous dynamical systems.

\subsection{Logistic map}\label{sec41}

\subsubsection{The $a=4$ case}\label{sec411}

\begin{figure*}
\begin{center}
\includegraphics[width=0.39\textwidth]{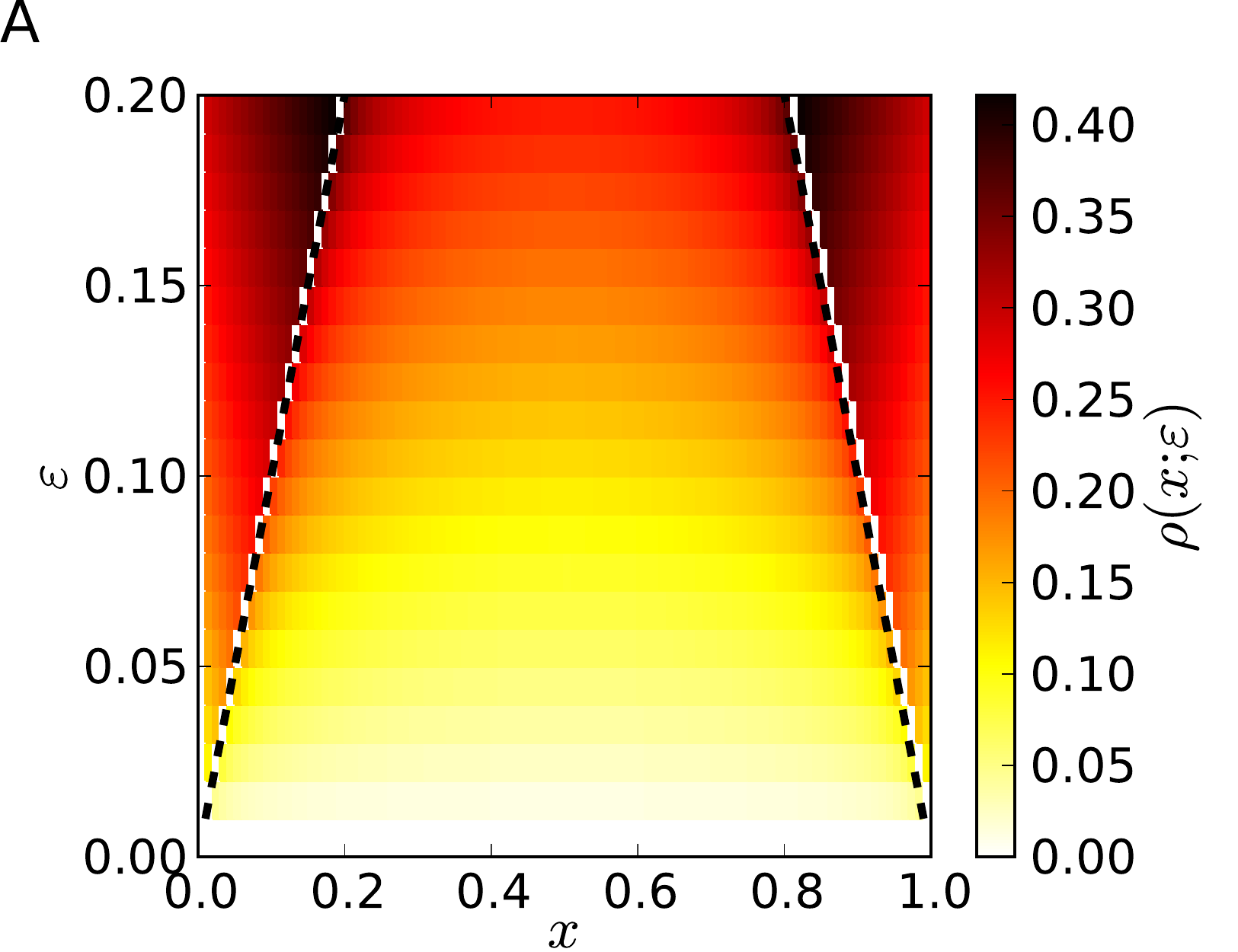} \hspace{0.5cm}
\includegraphics[width=0.39\textwidth]{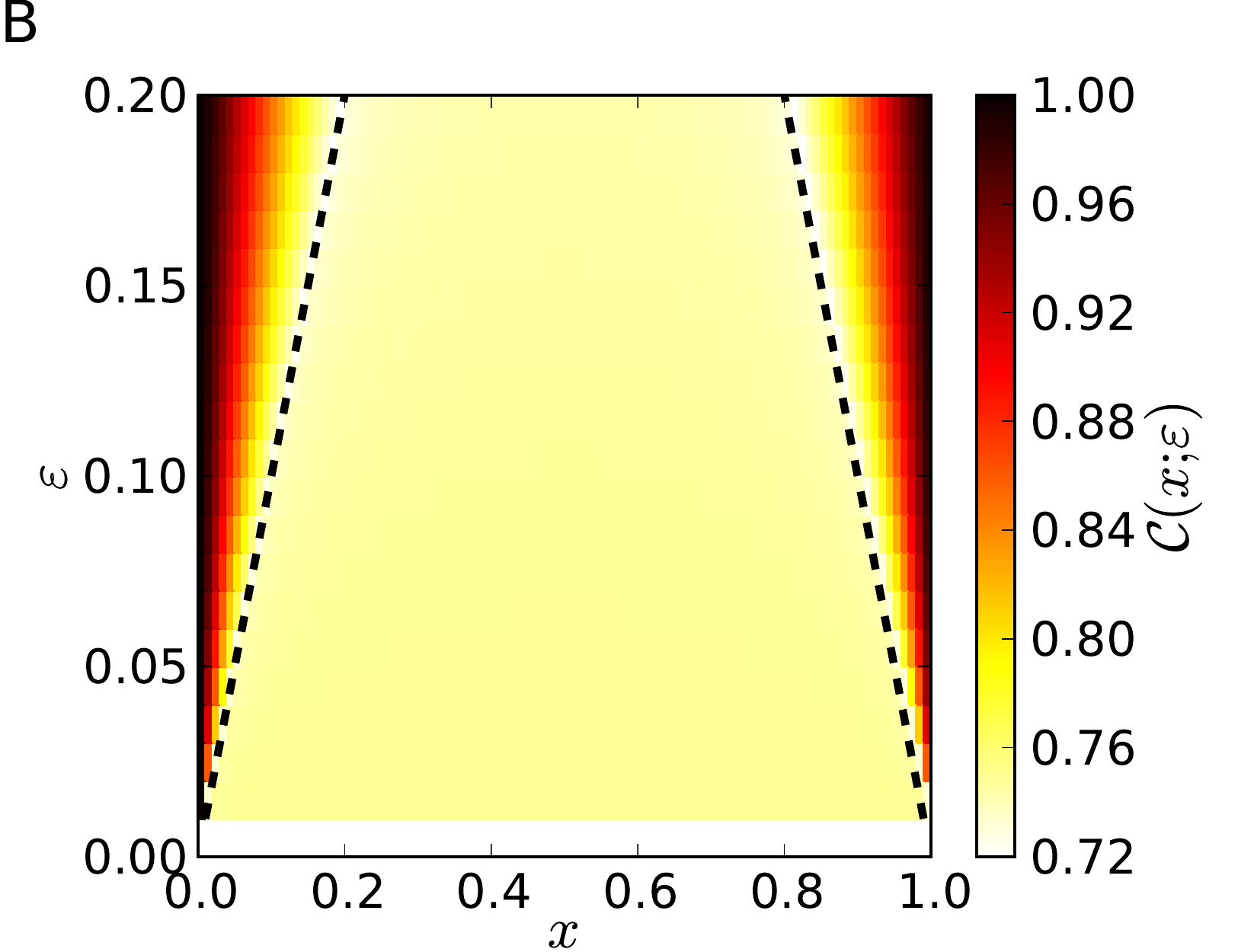} \\
\includegraphics[width=0.39\textwidth]{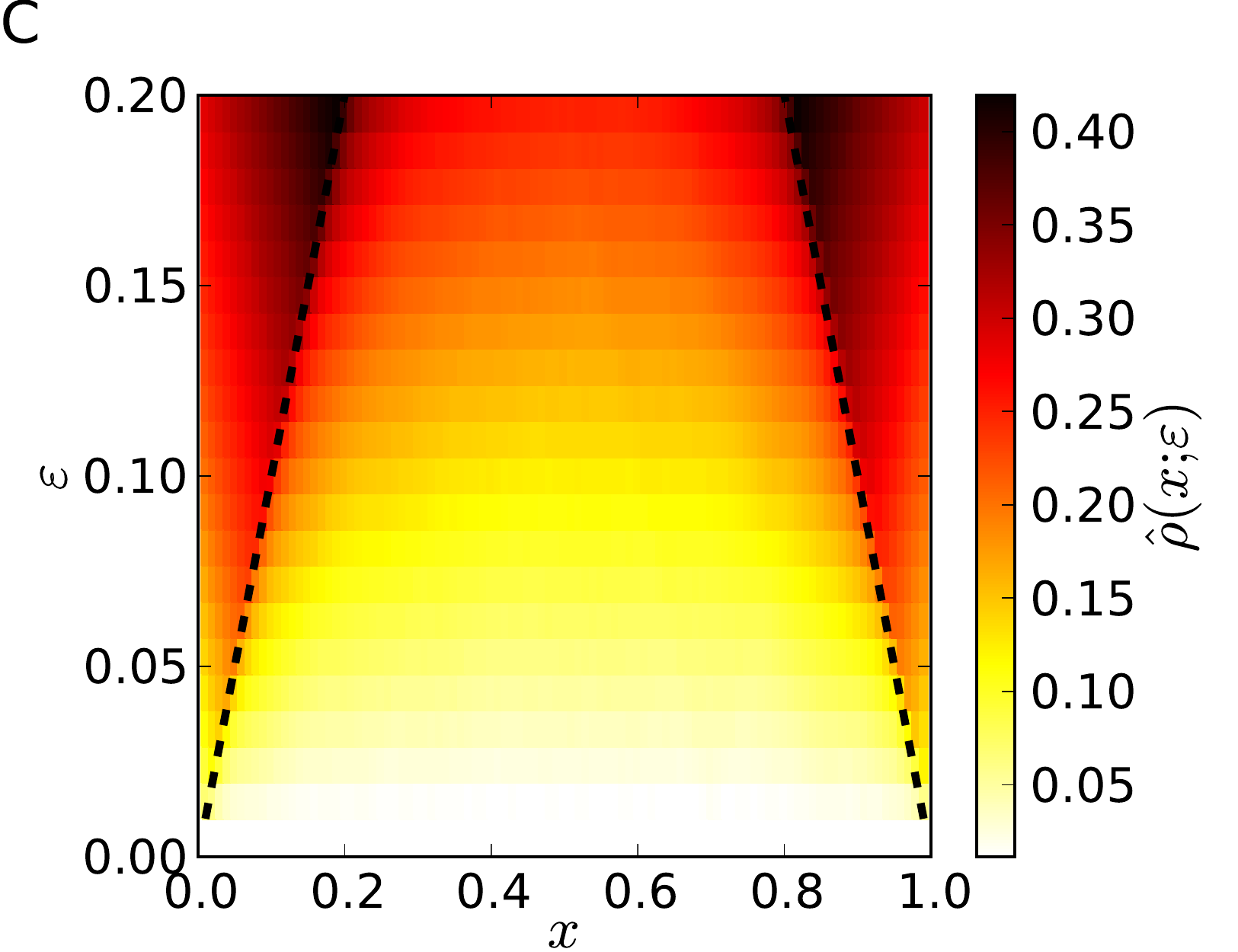} \hspace{0.5cm}
\includegraphics[width=0.39\textwidth]{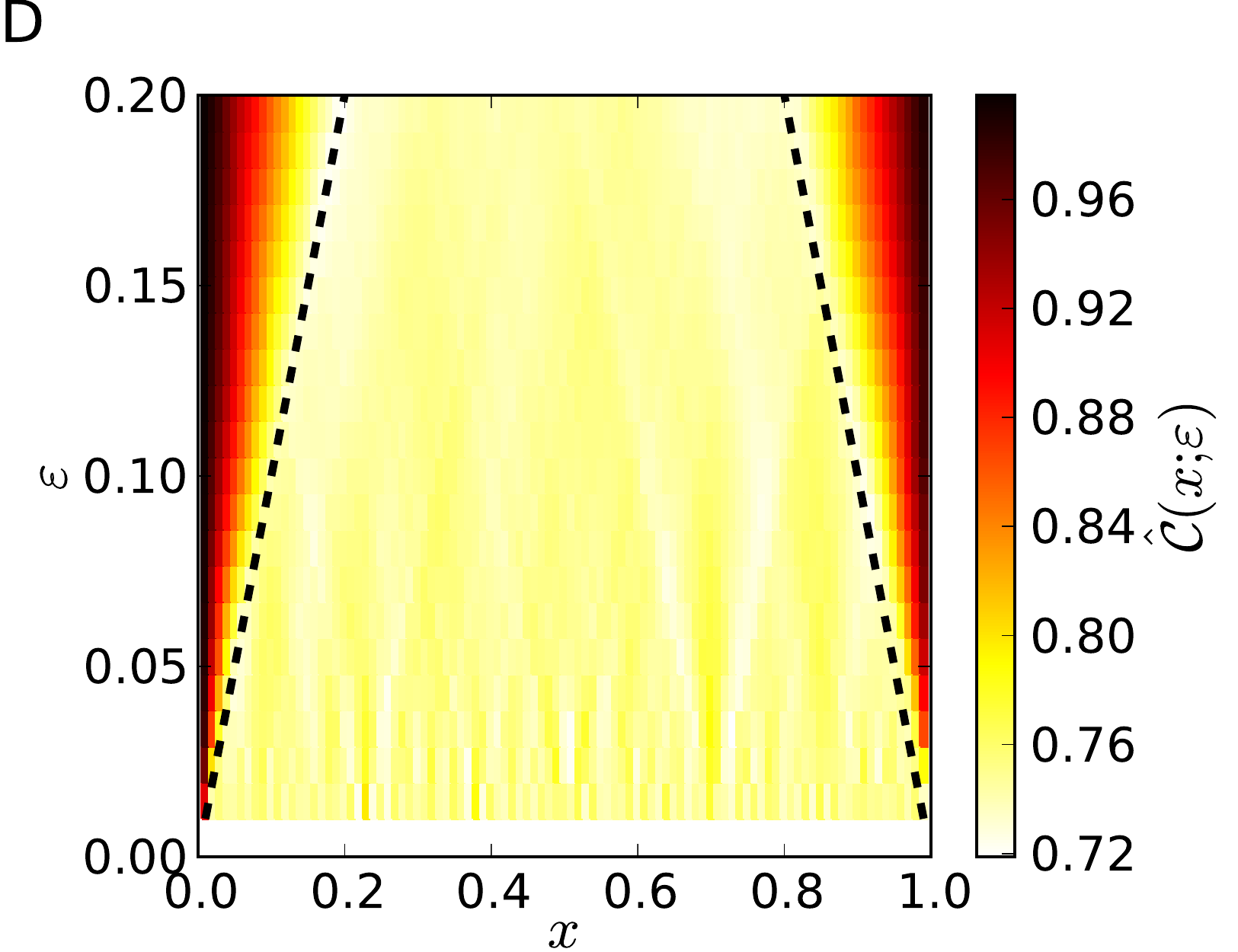}
\end{center}
\caption{Analytical values (upper panels) and numerical estimates (lower panels) of the local degree density $\rho(x;\varepsilon)$ (left) and the local clustering coefficient $\mathcal{C}(x;\varepsilon)$ (right) for the logistic map at $a=4$. Numerical results have been obtained for one realisation of the system with $N=10,000$ points. Diagonal lines indicate regions that are affected by information taken from outside the attractor.}
\label{fig:logmap_analytics}
\end{figure*}

For the logistic map (\ref{def:logmap}) at $a=4$, the knowledge of the invariant density $p(x)$ of the main attractor (Eq.~(\ref{logmap_density})) allows deriving analytical expressions for quantities such as local degree density and local clustering coefficients~\cite{Donner2010NJP}. In the latter case, one has to consider the simplification that the denominator in Eq.~(\ref{def:cloc}) factorises, i.e., the probabilities of two randomly chosen vertices $j$ and $k$ to be in the $\varepsilon$-ball around $x_i$ are independent and equal:
\begin{equation}
\begin{split}
P(A_{ij}=1,A_{ik}=1) & \simeq P(A_{ij}=1) P(A_{ik}=1)\\
& = P(A_{ij})^2 \approx \hat{\rho}_i^2.
\end{split}
\end{equation}

\begin{table*}[t]
\caption{Analytical expressions for the expectation values of the local degree densities and clustering coefficients for the logistic map at $a=4$ in dependence on both $x$ and $\varepsilon$~\cite{Donner2010NJP}. Note that the remaining integrals cannot be solved analytically.}

\noindent
Starting from the general identities
$$E[\rho(x,\varepsilon)]=\int_{x-\varepsilon}^{x+\varepsilon} p(y)\ dy$$
\noindent
and
$$E[\mathcal{C}(x,\varepsilon)]\approx\left[ \int_{x-\varepsilon}^{x+\varepsilon} dy\ p(y) \int_{\max(y-\varepsilon,x-\varepsilon)}^{\min(y+\varepsilon,x+\varepsilon)} dz\ p(z) \right]  \left[\int_{x-\varepsilon}^{x+\varepsilon} dy\ p(y) \right]^{-2},$$
we use the abbrevations 
\begin{eqnarray*}
I_1(a,b) &=& \int_a^b dy\ p(y) \\
I_2(a,b;c,d) &=& \int_a^b dy\ \left[ p(y) \int_c^d dz\ p(z) \right],
\end{eqnarray*}
\noindent
in order to obtain
$$E[\rho(x,\varepsilon)]=\left\{ \begin{array}{ll} 
I_1(0,x+\varepsilon), & \quad 0\leq x\leq\varepsilon, \\
I_1(x-\varepsilon,x+\varepsilon), & \quad \varepsilon\leq x\leq 1-\varepsilon, \\
I_1(x-\varepsilon,1), & \quad 1-\varepsilon\leq x\leq 1 \end{array} \right. $$
and
$$E[\mathcal{C}(x,\varepsilon)]\approx
\left\{ \begin{array}{ll}
I_1(0,x+\varepsilon)^{-2} \left(I_2(0,x;0,y+\varepsilon)+I_2(x,\varepsilon;0,x+\varepsilon)+I_2(\varepsilon,x+\varepsilon;y-\varepsilon,x+\varepsilon)\right), & \quad 0\leq x\leq\varepsilon, \\
I_1(x-\varepsilon,x+\varepsilon)^{-2}
\left(I_2(x-\varepsilon,x;x-\varepsilon,y+\varepsilon)+I_2(x,x+\varepsilon;y-\varepsilon,x+\varepsilon) \right), & \quad \varepsilon\leq x\leq 1-\varepsilon, \\
I_1(x-\varepsilon,1)^{-2}\left(I_2(x-\varepsilon,1-\varepsilon;x-\varepsilon,y+\varepsilon)+I_2(1-\varepsilon,x;x-\varepsilon,1)+I_2(x,1;y-\varepsilon,1)\right), & \quad 1-\varepsilon\leq x\leq 1. \end{array} \right. $$
\noindent
These expressions hold generally for one-dimensional maps defined on the unit interval. \\

\noindent
For the logistic map at $a=4$ with the invariant density $p(x)$ according to Eq.~(\ref{logmap_density}), one specifically finds
\begin{eqnarray*}
I_1(0,x+\varepsilon) &=& \frac{1}{2}-\frac{1}{\pi}\arcsin(1-2x-2\varepsilon), \\
I_1(x-\varepsilon,x+\varepsilon) &=& \frac{1}{\pi}\left[\arcsin(1-2x+2\varepsilon)-\arcsin(1-2x-2\varepsilon)\right], \\
I_1(x-\varepsilon,1) &=& \frac{1}{2}+\frac{1}{\pi}\arcsin(1-2x+2\varepsilon), \\
I_2(0,x;0,y+\varepsilon) &=& \frac{1}{4} - \frac{1}{2\pi}\arcsin(1-2x) - \frac{1}{\pi^2} \int_0^x dy\ \frac{\arcsin(1-2y-2\varepsilon)}{\sqrt{(y(1-y))}}, \\
I_2(x,\varepsilon;0,x+\varepsilon) &=& \left( \frac{1}{2\pi}-\frac{1}{\pi^2}\arcsin(1-2x-2\varepsilon) \right) \left( \arcsin(1-2x)-\arcsin(1-2\varepsilon) \right), \\
I_2(\varepsilon,x+\varepsilon;y-\varepsilon,x+\varepsilon) &=&
\frac{1}{\pi^2}\arcsin(1-2x-2\varepsilon)(\arcsin(1-2x-2\varepsilon)-\arcsin(1-2\varepsilon))+\frac{1}{\pi^2}\int_{\varepsilon}^{x+\varepsilon} dy\ \frac{\arcsin(1-2y+2\varepsilon)}{\sqrt{y(1-y)}}, \\
I_2(x-\varepsilon,x;x-\varepsilon,y+\varepsilon) &=&
\frac{1}{\pi^2}\arcsin(1-2x+2\varepsilon)(\arcsin(1-2x+2\varepsilon)-\arcsin(1-2x))-\frac{1}{\pi^2}\int_{x-\varepsilon}^{x} dy\ \frac{\arcsin(1-2y-2\varepsilon)}{\sqrt{y(1-y)}}, \\
I_2(x,x+\varepsilon;y-\varepsilon,x+\varepsilon) &=& \frac{1}{\pi^2}\arcsin(1-2x-2\varepsilon)(\arcsin(1-2x-2\varepsilon)-\arcsin(1-2x))+\frac{1}{\pi^2}\int_{x}^{x+\varepsilon} dy\ \frac{\arcsin(1-2y+2\varepsilon)}{\sqrt{y(1-y)}}, \\
I_2(x-\varepsilon,1-\varepsilon;x-\varepsilon,y+\varepsilon) &=& 
\frac{1}{\pi^2}\arcsin(1-2x+2\varepsilon)(\arcsin(1-2x+2\varepsilon)+\arcsin(1-2\varepsilon))-\frac{1}{\pi^2}\int_{x-\varepsilon}^{1-\varepsilon} dy\ \frac{\arcsin(1-2y-2\varepsilon)}{\sqrt{y(1-y)}}, \\
I_2(1-\varepsilon,x;x-\varepsilon,1) &=& -\left( \frac{1}{2\pi}+\frac{1}{\pi^2}\arcsin(1-2x+2\varepsilon) \right) \left( \arcsin(1-2x)+\arcsin(1-2\varepsilon) \right), \\
I_2(x,1;y-\varepsilon,1) &=& \frac{1}{4} + \frac{1}{2\pi}\arcsin(1-2x) + \frac{1}{\pi^2} \int_x^1 dy\ \frac{\arcsin(1-2y+2\varepsilon)}{\sqrt{(y(1-y))}}.
\end{eqnarray*}

\label{tab:logmap}
\end{table*}

Since the chaotic attractor is bound to the interval $(0,1)$, a careful treatment of the resulting integration boundaries reveals that spatial variations in all measures can be understood as being originated in the coverage of phase space regions outside the attractor (i.e., outside the interval $(0,1)$) rather than being a direct effect of the local degree density. The corresponding results obtained in \cite{Donner2010NJP} are briefly summarised in Tab.~\ref{tab:logmap}. The analytical considerations explain the numerical findings concerning the $\varepsilon$-de\-pen\-den\-ce of the global clustering coefficient $\mathcal{C}$ as well as the $x$-de\-pen\-den\-ce of the local clustering coefficient $\mathcal{C}_i$ for fixed $\varepsilon$ in an excellent manner~\cite{Donner2010NJP}. Figure~\ref{fig:logmap_analytics} demonstrates that this actually holds for both local degree density and clustering coefficient and for all $x$ and $\varepsilon$. 

\begin{figure}
\begin{center}
\includegraphics[width=0.48\textwidth]{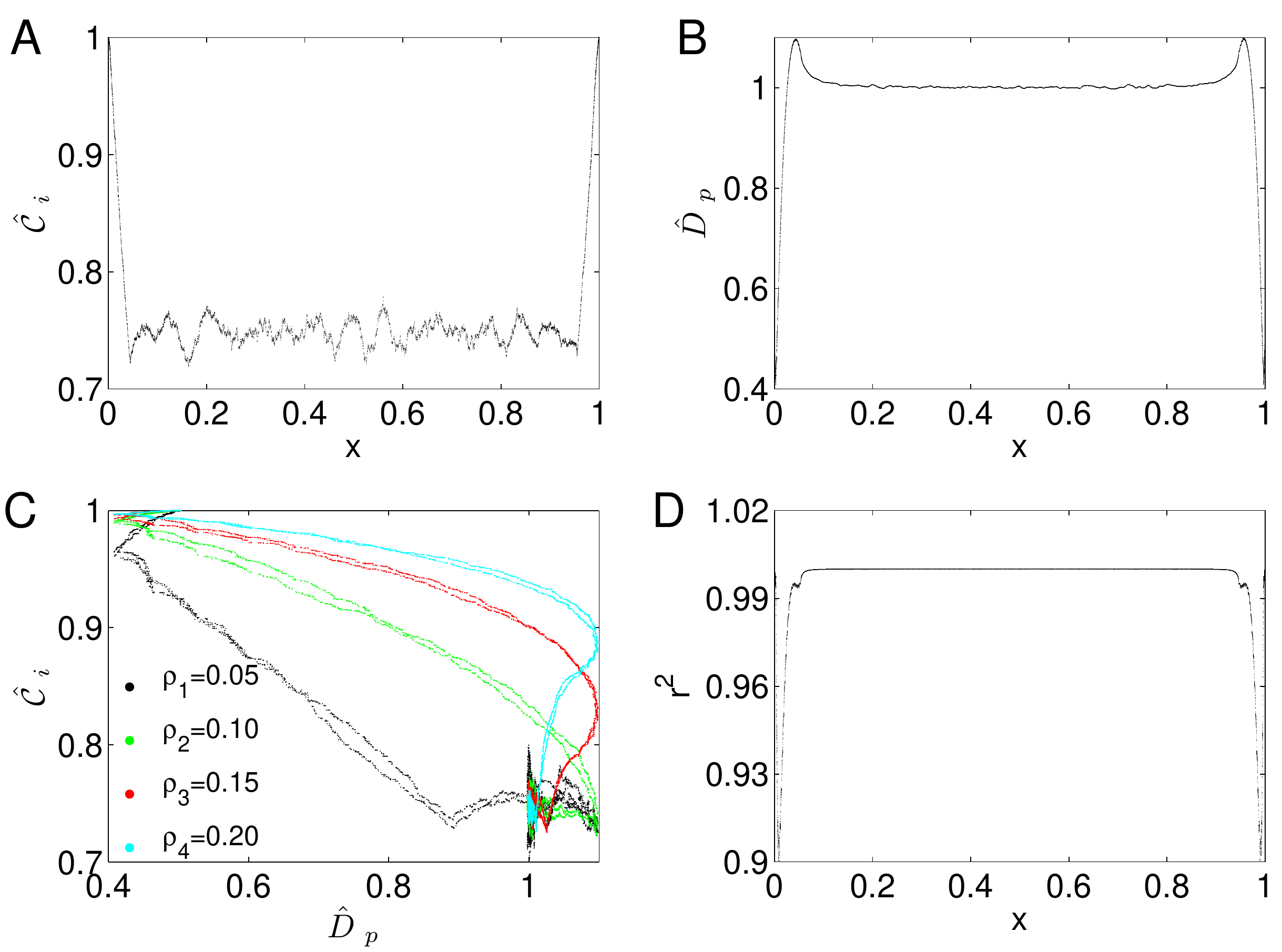}
\end{center}
\caption{Properties of the $\varepsilon$-recurrence networks obtained for one realisation of the logistic map at $a=4$: 
Point estimates of (A) $\hat{\mathcal{C}}_i(\varepsilon)$ and (B) $\hat{D}_p(x_i)$ as well as (D) the goodness-of-fit $r^2$ {of the linear regression of $\log\hat{\rho}_i$ vs.~$\log\varepsilon$ used} for the estimation of $\hat{D}_p(x_i)$ {(Eq.~(\ref{eq_dpest}))} in dependence on $x$. In addition, the relationship between estimates of the local clustering coefficient $\hat{\mathcal{C}}_i$ and pointwise dimension $\hat{D}_p(x_i)$ is shown for different choices of the edge density $\rho$ in panel (C). For the estimation of network properties, $N=10,000$ points have been used with an edge density of $\rho=0.01$, while estimates of the pointwise dimension have been obtained from $N=10^7$ points. The initial 100 points have been removed to avoid transient behaviour.}
\label{fig:logmap_density}
\end{figure}

From the dependence of the invariant density and, hence, the local degree density on both $x$ and $\varepsilon$, one may conclude that the pointwise scale-local dimension shows a similar dependence. However, when looking in more detail at estimates of the actual pointwise dimension $\hat{D}_p(x)$ obtained from the scaling exponent of the local degree density (Fig.~\ref{fig:logmap_density}), one finds that as expected, both $\hat{D}_p(x)$ and the scale-local (fixed $\varepsilon$) estimate $\hat{D}_{\mathcal{C},i}(\varepsilon)=\log\hat{\mathcal{C}}_i(\varepsilon)/\log(3/4)$ approach values close to 1 in a broad range in the middle of the chaotic attractor, i.e., in some interval that is not influenced by the attractor boundaries. The smaller $\varepsilon$ and the larger $N$, the better the pointwise convergence of $\hat{D}_p(x)\to 1$ and $\hat{D}_{\mathcal{C},i}(\varepsilon)\to 1$ obtained by numerical calculations for this region. We emphasise that the profile of the pointwise dimension is much smoother than that of the local clustering dimension, which follows from the fact that a variety of different values of $\varepsilon$ has been used in the estimation of $\hat{D}_p(x)$, while only one $\varepsilon$ had to be considered for $\hat{D}_{\mathcal{C},i}(\varepsilon)$, resulting in a larger variance of the estimate. 

Close to $x=\varepsilon$ and $1-\varepsilon$, the corresponding estimates become larger than the theoretical upper limit of $1$, whereas for $x\to 0,1$, $\hat{D}_p(x)\to 0$ and $\hat{D}_{\mathcal{C},i}(\varepsilon)\to 0$ as expected (see Sec.~\ref{sec321}). The observed overshooting of the estimates $\hat{D}_p(x)$ close to $x=\varepsilon,1-\varepsilon$ results from a loss of convergence of the estimator, which is underlined by Fig.~\ref{fig:logmap_density}D. Moreover, the scatter plot of $\hat{\mathcal{C}}_i$ versus $\hat{D}_p(x_i)$ for different $\varepsilon$ (Fig.~\ref{fig:logmap_density}A) demonstrates that apart from the regions close to the attractor boundaries, all points with an $\varepsilon$-neighbourhood that lies completely within $(0,1)$ are characterised by $\hat{D}_p(x_i)\approx\hat{D}_{\mathcal{C},i}(\varepsilon)\approx 1$ as expected. For the regions suffering from boundary effects (which are particularly pronounced in Fig.~\ref{fig:logmap_density}C due to rather large choices of $\varepsilon$, cf.~the discussion in~\cite{Donner2010PRE}), there is still a clear (but nonlinear) dependence between both measures. Note that the two-band structure in the scatter plot between both measures results from some numerical effects due to a slight asymmetry between the density close to the two attractor boundaries, which is expected to vanish for higher $N$. 

\subsubsection{Bifurcation scenario}\label{sec412}

It has already been shown that for the logistic map, regions with a high invariant density, which are typical for the supertrack functions, coincide with local maxima of both local degree density $\rho(x)$ and local clustering coefficient $\mathcal{C}(x)$ (Fig.~\ref{logmap_clustering}). For the local degree density, this observation is related to the fact that supertrack functions correspond to accumulation points of iterates of the map, in the vicinity of which trajectories tend to stay for a finite amount of time since they are only weakly repulsive in comparison with the usual exponential separation rate of the map~\cite{Lathrop1989}. A corresponding reasoning for $\mathcal{C}(x)$ has already been discussed in Sec.~\ref{sec321}.

\begin{figure}
\centering\includegraphics[width=0.48\textwidth]{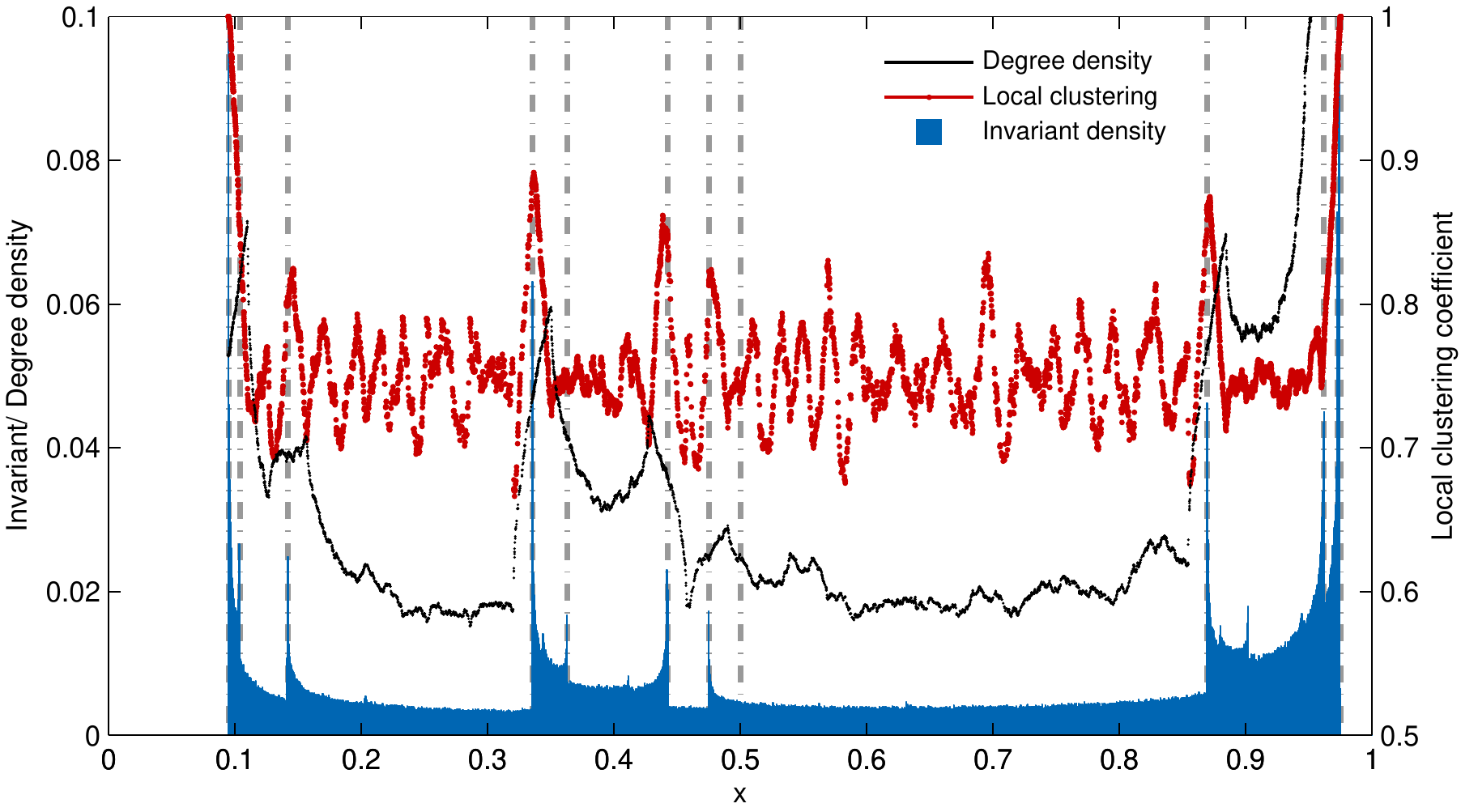}
\caption{Profile of {the invariant density $p(x)$ (blue) and estimates of the two network measures} $\hat{\rho}_i$ (black) and $\hat{\mathcal{C}}_i$ ({red}) for {the logistic map at} $a=3.9$ (parameters as in Fig.~\ref{logmap_clustering}). The positions of the first 12 supertrack functions are indicated by vertical lines. {Note that unlike the invariant density, the degree density is not a probability density, i.e., not normalised.}}
\label{logmap_supertrack}
\end{figure}

From the application point of view, one could ask whether $\rho(x)$ or $\mathcal{C}(x)$ are better suited for approximating the (possibly unknown) location of supertrack functions in a map. The specific supertrack shown in Fig.~\ref{logmap_clustering}D suggests that the local maximum of $\mathcal{C}(x)$ approximates the theoretical location better than that of $\rho(x)$. This observation is related to the argument from Sec.~\ref{sec321} that the invariant density $p(x)$ is approximately constant on one side of a supertrack function of the map, but decays like a power-law with increasing distance on its other side. As a consequence, estimating $\rho(x)$ with a finite $\varepsilon$ can be expected to result in a bias of the local maximum of $\rho(x)$. In order to study the generality of the latter finding, Fig.~\ref{logmap_supertrack} shows the complete profile of $\rho(x)$ and $\mathcal{C}(x)$ for $a=3.9$. One finds that at least the most pronounced interior maxima of the local clustering coefficient indeed coincide very well with supertrack functions of low order, where\-as the corresponding maxima of $\rho(x)$ are somewhat shifted from the known locations of the supertracks. We emphasise, however, that these shifts are {directly} related to our choice of $\epsilon$ (the same holds for the differences between the estimated $\hat{\mathcal{C}}_i$ and the theoretically predicted value of $1$ at the supertracks) and vanish in the limit $N\to\infty$, $\varepsilon\to 0$. For further higher-order supertrack functions, the {numerical} coincidence of maxima of $\mathcal{C}(x)$ with the {supertracks} is even weaker, which is also {a result of} the finite sample size and insufficient spatial resolution. Using longer realisations and smaller values of $\varepsilon$ {(not shown) yields} a more reliable profile and, hence, {improves} the skills of both vertex properties for localising supertrack functions.

\subsection{Two-dimensional maps}\label{sec42}

For the logistic map discussed above, the chaotic attractors often cover simply connected subintervals of $(0,1)$, with the possible exception of a countable set of isolated points on supertrack functions, which has measure $0$. In contrast to this case, there are numerous examples of chaotic maps that have attractors with a pronounced fractal structure. Following our theoretical considerations from Sec.~\ref{sec3}, it is interesting to study the behaviour of clustering and transitivity dimensions for such maps and compare it with the classical concepts of pointwise and local Lyapunov dimensions (see, e.g., \cite{Grebogi1988}).

\subsubsection{H\'enon map}\label{sec421}

As a first example, we consider the H\'enon map~\cite{Henon1976}
\begin{equation}
\begin{split}
x_{n+1} &= y_n+1-ax_n^2, \\
y_{n+1} &= bx_n,
\end{split}
\label{def:henonmap}
\end{equation}
\noindent
with the canonical parameters $a=1.4$ and $b=0.3$. The chaotic attractor of this map has a fractal structure, being smooth in one direction and a Cantor set in the other. Numerical estimates of the correlation dimension yield $D_2=1.42\pm 0.02$~\cite{grassberger83}. The pointwise dimension of the attractor has been extensively discussed in the framework of multifractal chaotic attractors and unstable periodic orbits~\cite{Grebogi1988}.

\begin{figure}
\begin{center}
\includegraphics[width=0.48\textwidth]{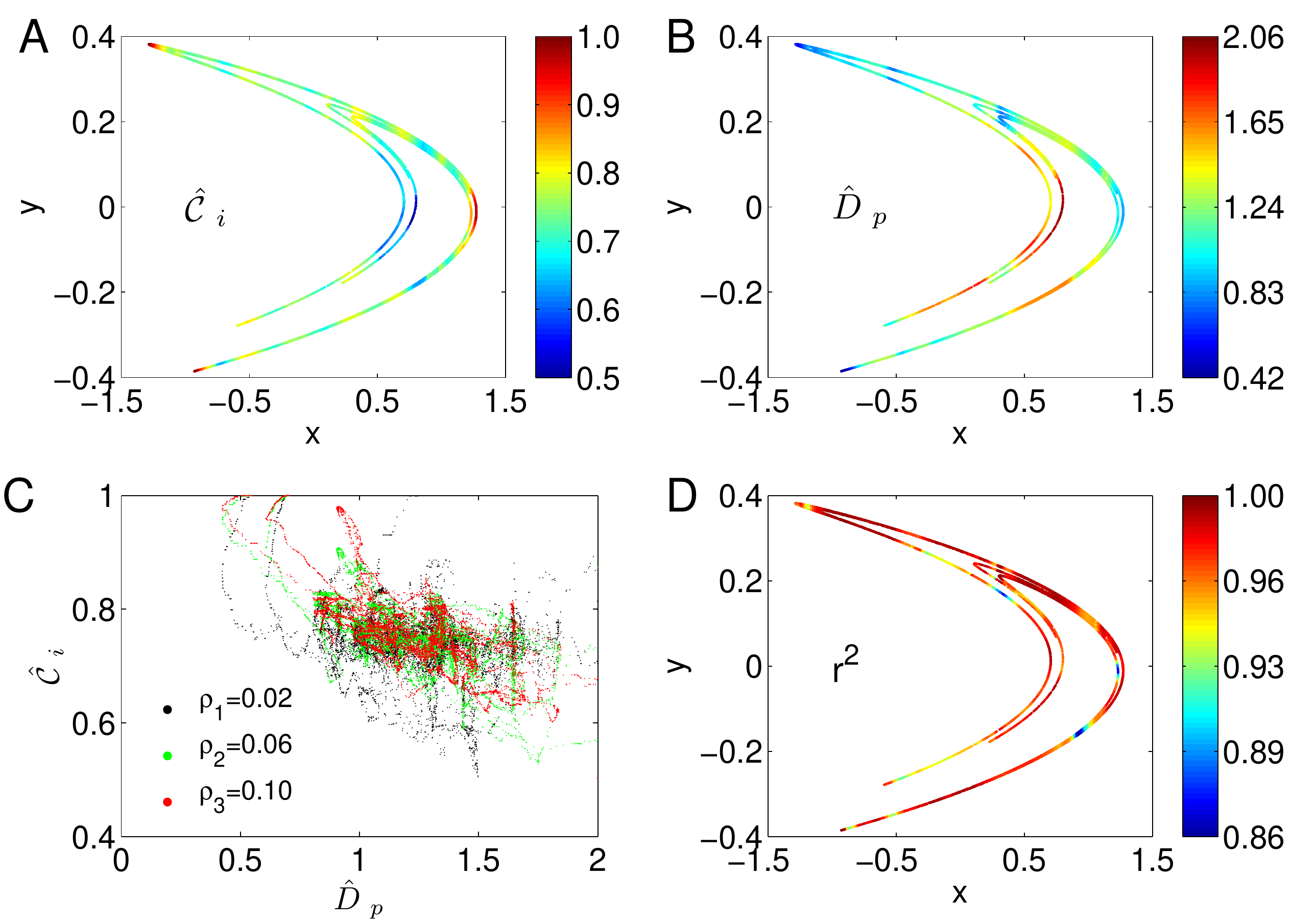}
\end{center}
\caption{Properties of the $\varepsilon$-recurrence networks obtained for one realisation of the H\'enon map at $a=1.4$ and $b=0.3$: Colour-coded representation of point estimates of (A) $\hat{\mathcal{C}}_i$ and (B) $\hat{D}_p(x_i)$ as well as (D) the goodness-of-fit $r^2$ for the estimation of $\hat{D}_p(x_i)$ in dependence on $x$ and $y$. In addition, the relationship between estimates of the local clustering coefficient $\hat{\mathcal{C}}_i$ and pointwise dimension $\hat{D}_p(x_i)$ is shown for different choices of the edge density $\rho$ in panel (C). For the estimation of network properties, $N=10,000$ points have been used with an edge density of $\rho=0.01$, while estimates of the pointwise dimension have been obtained from $N=10^6$ points. The initial 1000 points have been removed to avoid transient behaviour.}
\label{fig:henon_density}
\end{figure}

Figure~\ref{fig:henon_density} shows colour-coded representations of the local clustering coefficients and pointwise dimensions. Unlike for the logistic map, we find no clear relationship between both measures. A pronounced exception are the tips of the attractor, which locally represent zero-dimensional structures ($D_\mathcal{C}(x,y)\to 0$, $D_p(x,y)\to 0$). The strong differences between the respective measures, which can be found in large parts of the attractor, seem to be a consequence of the specific filamental structure of the attractor. In fact, the H\'enon attractor has a fractal support, which results in rather specific topological and metric properties~\cite{Cvitanovic1988}. 

\begin{figure}
\begin{center}
\includegraphics[width=0.4\textwidth]{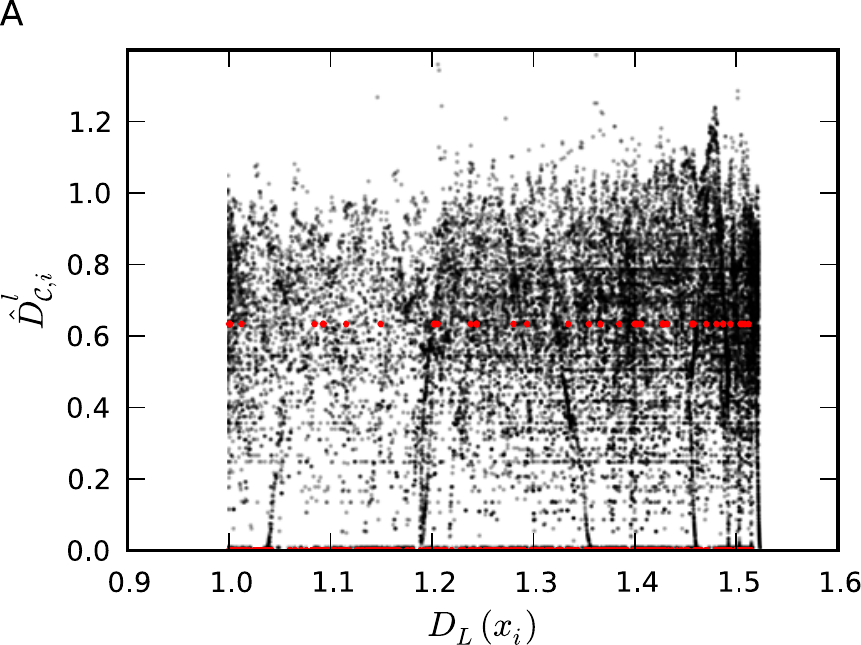} \\
\includegraphics[width=0.4\textwidth]{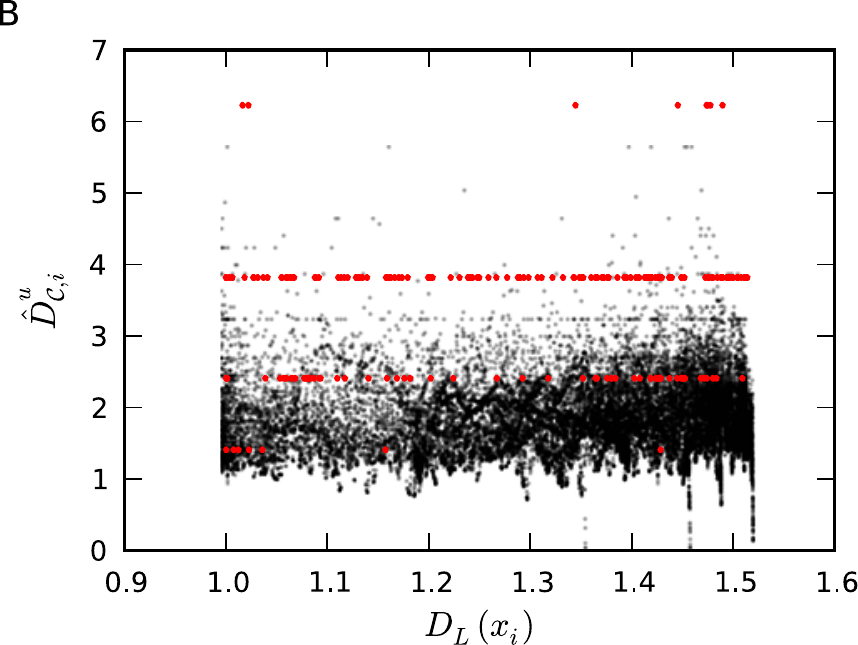}
\end{center}
\caption{Relationships between point estimates of (A) lower and (B) upper clustering dimensions on the one hand, and local Lyapunov dimensions on the other hand, for one realisation of the H\'enon map at $a=1.4$ and $b=0.3$ estimated from $N=30,000$ data points (initial condition $(x,y)=(0,0)$, the first $1,000$ iterations have been removed from the trajectory to avoid transient behaviour) using 100 equally spaced values of $\varepsilon$ in the interval $[0.005,0.1]$. Red dots correspond to vertices with low degree $k_i<5$, for which the obtained estimates of $\hat{D}_{\mathcal{C},i}^{u,l}$ are hardly significant.}
\label{fig:henon_cdim}
\end{figure}

A similar inconsistency can be observed when comparing the upper and lower clustering dimensions with the local Lyapunov dimension (see Fig.~\ref{fig:henon_cdim}), where no clear statistical relationship seems to exist as well. Nevertheless, the local clustering dimensions behave in the expected way: the upper clustering dimension $\hat{D}_{\mathcal{C}}^u$ is smaller than the upper bound $2.41m\approx 4.82$ discussed in Sec.~\ref{sec323} for almost all vertices, with only very few exceptions corresponding to vertices with low degree (Fig.~\ref{fig:henon_cdim}B). A similar observation is made for $D_\mathcal{C}^l$, which is always smaller than the dimension of the surrounding space ($m=2$) and shows non-zero values for the vast majority of vertices (Fig.~\ref{fig:henon_cdim}A). Most vertices with zero values of $\hat{D}_{\mathcal{C},i}^l$ have very low degree as well, pointing to a purely statistical effect. However, there are some exceptions such as vertices with some very specific location, e.g., close to the tips of the attractor bands.

\begin{figure}
\begin{center}
\includegraphics[width=0.4\textwidth]{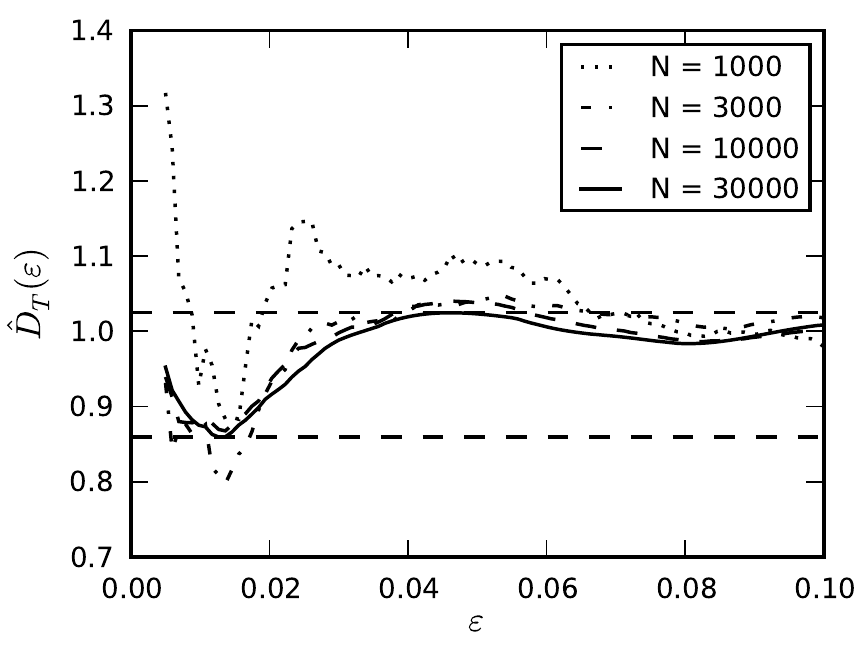}
\end{center}
\caption{Estimation of the transitivity dimensions $\hat{D}_{\mathcal{T}}^{u,l}$ of the H\'enon map at $a=1.4$ and $b=0.3$ (one realisation with initial condition $(x,y)=(0,0)$, the first $1,000$ iterations have been removed from the trajectory to avoid transient behaviour) for different $N$ obtained with the same set of thresholds $\varepsilon$ as in Fig.~\ref{fig:henon_cdim}. Dashed horizontal lines indicate numerical estimates of $D_\mathcal{T}^{u,l}$ obtained with $N=30,000$ data points.}
\label{fig:henon_transdim}
\end{figure}

As a consequence of the aforementioned observations, very long realisations are typically required to numerically capture the local features of the chaotic attractor of the H\'enon map with reasonable confidence. This is further underlined by the transitivity dimensions: the larger $N$, the better the estimated values of this measure obtained for fixed $\varepsilon$ approach stationary values corresponding to the upper and lower transitivity dimensions (Fig.~\ref{fig:henon_transdim}). In contrast, for too small $N$, we observe significant deviations from the asymptotically estimated values, which becomes particularly important for $\varepsilon\to 0$.

\subsubsection{Generalised baker's map}\label{sec422}

\begin{figure}
\begin{center}
\includegraphics[width=0.48\textwidth]{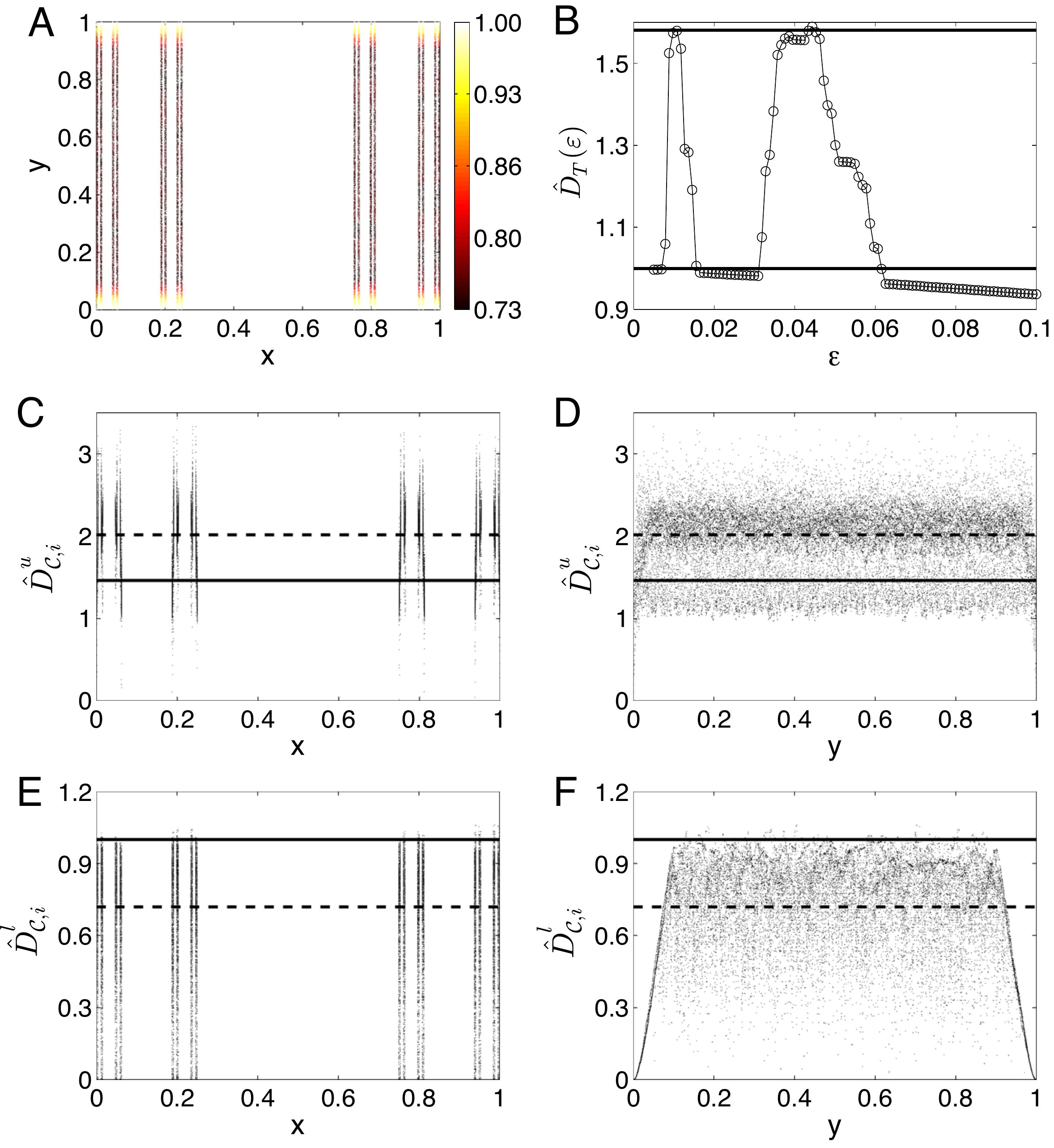}
\end{center}
\caption{Estimates of the clustering dimensions obtained for one realisation of the standard baker's map ($\alpha=1/2$ and $\lambda_a=\lambda_b=1/4$) with initial conditions $(x_0,y_0)=(0.2,0.2)$. All results have been obtained with $N=30,000$ data points. Initial transients have been avoided by removing the first $1,000$ iterations from the sample trajectory. (A) Colour-coded representation of the local clustering coefficient $\hat{\mathcal{C}}_i$ obtained with $\varepsilon=0.1$. (B) Scale-local estimates of the transitivity dimension $\hat{D}_{\mathcal{T}}(\varepsilon)$ in dependence on the considered $\varepsilon$. Horizontal lines indicate the true values of these measures derived analytically. Qualitatively the same results have been obtained for other (lower) choices of $N$ (not shown here), indicating reasonable convergence properties of the proposed estimator. In addition, point estimates of the local upper (C,D) and lower (E,F) clustering dimensions $\hat{D}_{\mathcal{C},i}^{u,l}$ (estimated with 100 equally spaced values of $\varepsilon$ in $[0.005,0.1]$) are shown in dependence on both variables $x$ (C,E) and $y$ (D,F). Solid horizontal lines correspond to the theoretical values, whereas dashed lines indicate the median values obtained from the considered sample of state vectors.}
\label{fig:gbm_cdim}
\end{figure}

For the symmetric version of the baker's map (Eq.~\ref{genbakersmap}), one finds that for $y\neq 1/2$, the Jacobian is given as $\mathop{\rm diag}(1/4,2)$ with singular values $a_1=2\geq a_2=1/4$. Hence, the local Lyapunov dimension is $D_L(x,y)=1+1/2$ since $a_1\geq 1>a_1a_2$ and $a_1a_2^{1/2}=1$. For the clustering and transitivity dimensions, we have shown in Sec.~\ref{sec322} that for almost all $(x,y)\in S$, $D_\mathcal{C}^u(x,y) \gtrsim 1.464$, $D_\mathcal{T}^u\approx 1.581$, and $D_\mathcal{C}^l(x,y)=D_\mathcal{T}^l=1$. The latter results are confirmed by numerical calculations, the results of which are summarised in Fig.~\ref{fig:gbm_cdim}. It is notable that the estimated values of the transitivity dimension roughly coincide with the theoretically predicted upper and lower bounds. However, there are examples where these bounds are exceeded. We identify two possible reasons for such behaviour: too large $\varepsilon$ or (for small $\varepsilon$) too small $N$, i.e., finite-scale and finite sample size effects. Due to the resulting outliers obtained when varying $\varepsilon$, numerical values for the upper (lower) clustering dimension typically have a positive (negative) bias with respect to the theoretically predicted values, which is nicely illustrated by Fig.~\ref{fig:gbm_cdim}C-F. As a statistical estimation effect, this bias is more severe for local dimensions, since the variance is larger than for transitivity dimensions (see above). However, a bias also exists for the transitivity dimensions, where it is just smaller (e.g., see the overshooting in Fig.~\ref{fig:gbm_cdim}B).

For the generalised baker's map, detailed analytical expressions are available for the global Lyapunov dimension of the system~\cite{Farmer1983,Ott1993}. For the local version of this measure, we obtain the following results:
\begin{equation}
D_L(x,y) = \begin{cases} 1 + \frac{\ln \alpha}{\ln \lambda_a} & , \, y < \alpha \wedge \alpha > \lambda_a \\
		2 & , \, y < \alpha \wedge \alpha \leq \lambda_a \\ 
		1 + \frac{\ln (1-\alpha)}{\ln \lambda_b} & , \, y > \alpha \wedge 1-\alpha > \lambda_b \\
		2 & , \, y > \alpha \wedge 1-\alpha \leq \lambda_b.  \end{cases}
\label{dl_gbm}
\end{equation}
\noindent
These expressions can be used as a benchmark with which we can compare the numerical estimates of our new measures $\hat{D}_{\mathcal{C},i}^{u,l}$ and $\hat{D}_\mathcal{T}^{u,l}$. 

In contrast to the Lyapunov dimensions, simple expressions for the local clustering and transitivity dimensions of the generalised baker's map can (unlike for its symmetric version) only be obtained for some specific cases. For example, concerning the dependence on $\alpha$ for fixed $\lambda_a=\lambda_b=1/4$ and $\varepsilon=7/4^n$ ($n>1$), the transitivity can be calculated as 
\begin{equation}
\mathcal{T} = 1- \frac{(2\alpha^2(1-\alpha)^2)(\alpha^2 + (1-\alpha)^2)}{1 + \alpha(1-\alpha)(2\alpha(1-\alpha)(\alpha(1-\alpha)+1)-3)},
\label{gbmt}
\end{equation}
\noindent
which allows deriving a corresponding expression for $D_\mathcal{T}$. In the derivation of the latter expression, we have used the fact that each linked triple lies in a small band of width $16/4^n$ that is composed of four substrips of width $1/4^n$ with relative weights of $\alpha^2$, $\alpha(1-\alpha)$, $\alpha(1-\alpha)$, and $(1-\alpha)^2$, and gaps of width $1/4^n$, $5/4^n$, and $1/4^n$, respectively. For other values of $\varepsilon$, the transitivity might be even smaller, leading to larger values of $D_{\mathcal{T}}$. Hence, the estimate based on Eq.~(\ref{gbmt}) has to be considered a lower bound for the actual value of the upper transitivity dimension $D_\mathcal{T}^u$.

\begin{figure}
\begin{center}
\includegraphics[width=0.4\textwidth]{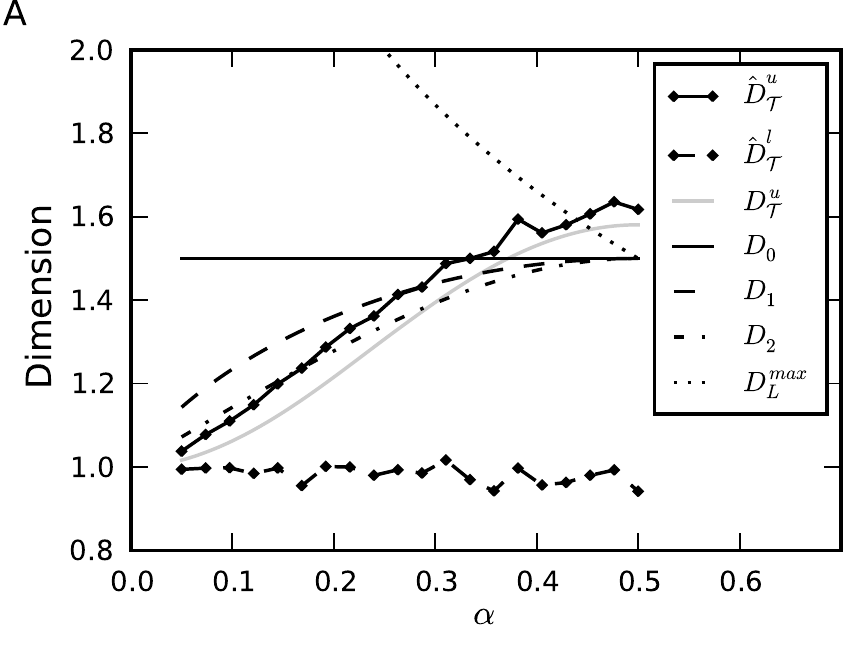} \\
\includegraphics[width=0.4\textwidth]{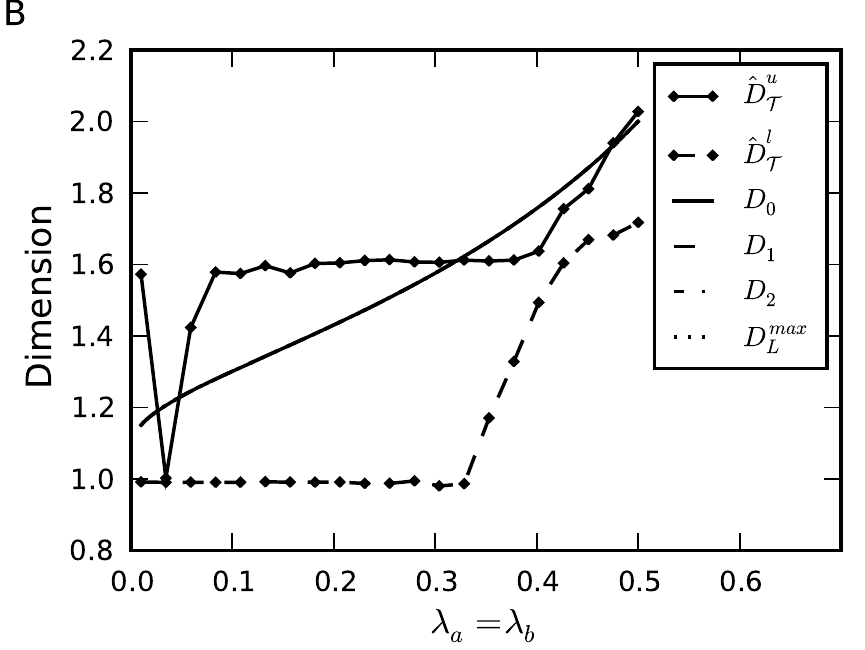}
\end{center}
\caption{Dependence of several global measures of dimensionality on the parameters (A) $\alpha$ ($\lambda_a=\lambda_b=1/4$) and (B) $\lambda_a=\lambda_b$ ($\alpha=1/2$) of the generalised baker's map (here, one realisation has been considered for each parameter combination, initial conditions and removal of initial transients as in Fig.~\ref{fig:gbm_cdim}). The grey line in (A) corresponds to the theoretical lower bound of $D_\mathcal{T}^u$ (see text). Note that in (B), all ``classical'' measures ($D_0$, $D_1$, $D_2$ and $D_L^{max}$) have equal values. Numerical estimates have been obtained with  $N=15,000$ (A) and $N=25,000$ (B) data points, respectively, using $50$ equidistant values of $\varepsilon\in[0.001,0.015]$ (in this range, $\hat{D}_\mathcal{T}^u$ and $\hat{D}_\mathcal{T}^l$ approximate the analytical results well for the $\alpha=0.5$ case as shown in Fig.~\ref{fig:gbm_cdim}B). The results are robust for various choices of N.}
\label{fig:gbm_globdim}
\end{figure}

Figure~\ref{fig:gbm_globdim}A shows the $\alpha$-dependence for different notions of dimension. Analytical results for the ``classical'' measures $D_0$, $D_1$ and $D_2$ have been taken from~\cite{Ott1993}. Note that due to its definition~\cite{Hunt1996}, the maximum local Lyapunov dimension $D_L^{max}$ is bound from above by the dimension of the underlying phase space ($m=2$, see Eq.~(\ref{dl_gbm})), although numerical calculations would yield higher values for $\alpha<\lambda_a$. We emphasise that setting some dimension of a given set equal to that of the surrounding space ($m$) whenever its numerical value exceeds $m$ avoids pathological behaviour, which has been observed in a similar way for the clustering and transitivity dimensions introduced in this paper (see Sec.~\ref{sec323}). Concerning the numerically estimated transitivity dimensions, we observe that $\hat{D}_\mathcal{T}^u$ coincides rather well with the lower analytical bound resulting from (\ref{gbmt}), but shows some positive bias. There are two possible reasons for this: (i) either the estimate (\ref{gbmt}) for the lower bound of the transitivity $\mathcal{T}$ is still too conservative, or (ii) the numerical values are too large due to some overshooting as indicated in Fig.~\ref{fig:gbm_cdim}B. Comparing $\hat{D}_\mathcal{T}^{u,l}$ with the other dimension measures, it becomes evident that in general, the transitivity dimension is neither an upper nor a lower bound to any of the considered classical concepts.

The dependence of different dimension measures on $\lambda_a=\lambda_b$ is shown for $\alpha=1/2$ in Fig.~\ref{fig:gbm_globdim}B. Note that for this specific choice of $\alpha$, all ``classical'' dimensions take the same values, which only depend on $\lambda_a=\lambda_b$. For sufficiently small $\lambda_a=\lambda_b\lesssim 0.3$, the upper and lower transitivity dimensions seem to take stationary values, which is in contrast to the other measures of dimensionality. For $\lambda_a=\lambda_b\to 0.5$, the map fills the complete two-dimensional unit box, so that all dimensions converge to 2 (note that this limit is not approached by the lower transitivity dimension due to the finite length of the considered realisation). For too small $\lambda_a=\lambda_b$, the numerical behaviour also suggests that longer realisations of the system are necessary to obtain reasonable results. In general, we conclude that for the parameter range within which our results can be considered reliably, the numerically estimated transitivity dimensions take similar values as the other measures and show a similar behaviour if the parameters of the generalised baker's map are varied (with the exception of the maximum local Lyapunov dimension), which suggests that the new net\-work-based dimensions are reasonably defined.

In all cases, note that the numerically estimated values of $\hat{D}_\mathcal{T}^{u,l}$ show some residual variations superimposed to their general trend. Besides the finite $N$, this is mainly due to the fact that only one specific realisation of the system at every set of parameters is used. We expect results to further improve if mean values taken from ensembles of independent realisations are considered.

\subsection{R\"ossler system}\label{sec43}

So far, we have only discussed examples of discrete maps. Among the dynamical systems showing complex behaviour, there are however many examples that are time-con\-ti\-nuous rather than discrete. In the following, we will discuss as one paradigmatic example the well-studied R\"ossler system
\begin{equation}
\begin{split}
\dot{x} &= -y-z \\
\dot{y} &= x+ay \\
\dot{z} &= b+z(x-c)
\end{split}
\end{equation}
\noindent
with the parameters $a=0.2$, $b=0.2$ and $c=5.7$. For the latter choice, the R\"ossler system is known to have a chaotic attractor. Moreover, there are countably many unstable periodic orbits (UPOs) of various periods, {which do not belong to the attractor, but are densely embedded in it and support the invariant measure. Hence, these} UPOs form a subset of the attractor's closure, which has measure 0. As a consequence, traditional dimension estimates typically characterise the properties of the cha\-otic part, but are not suited for describing the properties of the embedded unstable, but dynamically invariant periodic structures. In contrast, our results obtained for the logistic map suggest that local transitivity properties of $\varepsilon$-recurrence networks can be used for identifying at least the least repulsive UPOs. In the following, we will further discuss this idea and present some numerical results using the concept of continuous clustering and transitivity dimensions introduced in this paper.

\subsubsection{General considerations}\label{sec431}

Generalising the previous considerations concerning the supertrack functions of the logistic map (see Sec.\ \ref{sec321} and \ref{sec412}) to time-continuous dynamical systems, it appears a reasonable assumption that the continuous $\varepsilon$-clustering coefficient $\mathcal{C}(x;\varepsilon)$ is in general a consequence of the spatial alignment of neighbouring trajectories in phase space, which is closely related to the effective local dimension of the attractor. In this respect, we note that continuous systems may be transformed into discrete maps by choosing a proper Poin\-ca\-r\'e section. For example, supertrack-like structures in Poin\-ca\-r\'e sections of the Lorenz system can be identified using the vertex properties (in particular, degree and local clustering coefficient) of the associated $\varepsilon$-re\-cur\-ren\-ce networks~\cite{Donner2010IJBC}.


Besides our specific considerations for maps, we no\-te that in general, spatial differences in $\mathcal{C}(x;\varepsilon)$ (and, hence, $D_\mathcal{C}^{u,l}(x)$) can be theoretically understood using results for random geometric graphs~\cite{Dall2002}. Recall that since individual recurrence points are assumed to be separated by sufficiently large time intervals (i.e., sojourn points are excluded)~\cite{Donner2010NJP}, the actual spatial location of the associated vertices depends on the specific sampling of the data. Therefore, an $\varepsilon$-recurrence network can be interpreted as a random geometric graph with a certain effective dimension (in our case characterised by $D_\mathcal{T}$). We note that the latter considerations apply both globally and locally, i.e., they also hold for arbitrary subgraphs of an $\varepsilon$-re\-cur\-ren\-ce network. Since for arbitrary geometric graphs, the subgraph properties follow from the spatial distribution of vertices, spatial heterogeneities in this distribution can result in (among others) different local transitivity properties and, hence, a non-trivial spatial pattern of the pointwise (scale-local) clustering dimension. 

We emphasise that a low local dimension ($<m$) of the attractor implies that trajectories cannot (locally) exponentially diverge in all directions of the $m$-di\-men\-sio\-nal phase space, but rather become (locally) almost parallel in some lower-dimensional subspace. Among other cases, the latter behaviour can be considered typical in the vicinity of UPOs, where trajectories become dynamically trapped near an invariant lower-di\-men\-sio\-nal object for a certain finite time~\cite{Lathrop1989}. Since for random geometric graphs, it is known~\cite{Dall2002} (and verified by our analytical considerations in Sec.~\ref{sec3}) that the expected clustering coefficient decreases roughly exponentially with increasing spatial dimension of such networks, the hypothesis that $\mathcal{C}(x;\varepsilon)$ takes local maxima close to UPOs appears justified, which translates into a low pointwise scale-local clustering dimension $D_\mathcal{C}(x;\varepsilon)$. From this perspective, $D_\mathcal{C}$ directly relates to traditional concepts like pointwise dimensions (which, however, would typically characterise the chaotic attractor rather than the embedded UPOs) and local Lyapunov dimensions (which has, however, only been formally defined for maps so far). Moreover, we note that there is a direct link between Lyapunov dimension and Lyapunov exponents, which measure the average divergence rate of neighbouring trajectories and can be used for a local attractor characterisation as well (see definition).



\subsubsection{Period-3 UPOs}\label{sec432}

\begin{figure}
\begin{center}
\includegraphics[width=0.48\textwidth]{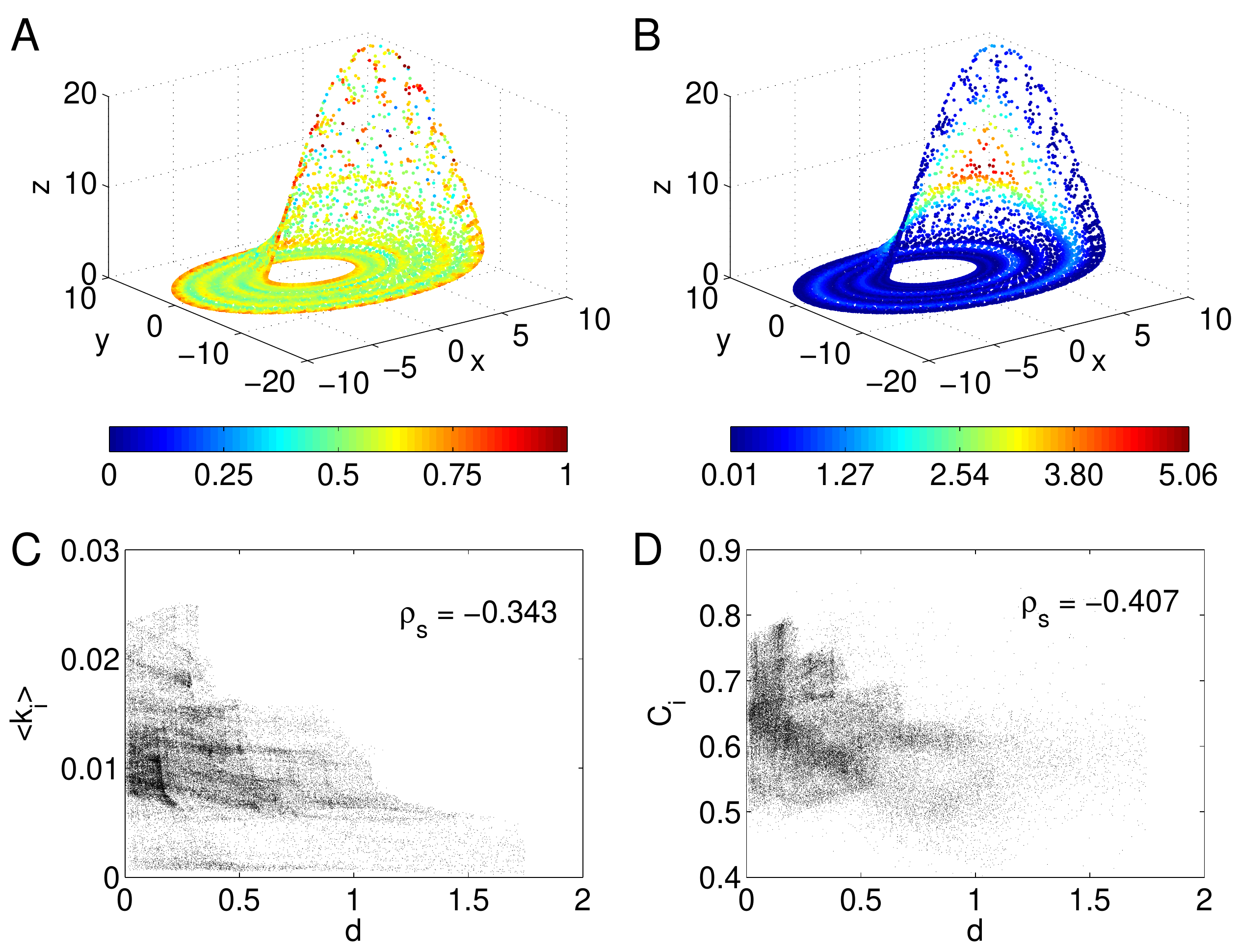}
\end{center}
\caption{Point estimates of (A) the local clustering coefficient $\hat{\mathcal{C}}_i$ and (B) the minimum Euclidean distance $d_i$ from the period-3 UPOs of the R\"ossler system ($N=10,000$, $\rho=0.01$). In addition, the dependence of (C) degree density and (D) local clustering coefficient on $d_i$ are shown for short distances ($N=50,000$, $\rho=0.01$). We emphasise that $d$ has been measured here using the Euclidean norm, whereas for the generation of the $\varepsilon$-recurrence networks, the maximum norm has been considered. Note that the respective (rank-order) correlations $\rho_S$ between vertex properties and distance from the UPO are significant and of comparable order for both measures.}
\label{fig:upos}
\end{figure}

As an empirical verification of the above consideration, we consider the dependence of the local clustering coefficient $\hat{\mathcal{C}}_i(\varepsilon)$ on the spatial coordinates of a vertex. Specifically, we study the distance of vertices from the two period-3 UPOs embedded in the chaotic R\"ossler attractor, which are particularly well expressed features of the system. As it follows from Fig.~\ref{fig:upos}, there is a clear indication that close to these UPOs, both vertex degree and local clustering coefficient show increased values (note that due to the three-dimensionality of the system, these maxima are not as well expressed as in the case of, e.g., the logistic map, particularly for some finite $\varepsilon$ smearing out the spatial signatures of the UPOs). At somewhat larger distances from these invariant objects, we find a clear tendency towards smal\-ler values of both measures indicated by significant negative values of the rank-order correlation co\-ef\-fi\-cients $\rho_S$. For the degree, this is clearly a consequence of the trapping feature of UPOs~\cite{Lathrop1989}, while according to our theoretical considerations, the corresponding result for the local clustering coefficient (and, hence, the associated local clustering dimension) is caused by the low dimensionality of the UPOs in comparison to the chao\-tic attractor itself.

For UPOs of higher periods (recall that these are densely embedded in the chaotic attractors), we however find much weaker signatures in the spatial pattern of $\mathcal{C}(x)$~\cite{Donner2010NJP}. This implies that the detection of high-periodic UPOs (which are typically more repulsive and, hence, characterised by shorter residence ti\-mes in their direct vicinity than UPOs of lower period) by means of $\varepsilon$-recurrence networks probably requires lon\-ger time series (larger $N$) and lower recurrence thresholds $\varepsilon$. We note that in principle, UPOs can also be detected by other types of proximity-based complex network approaches to time series analysis, for example, cycle networks~\cite{Zhang2006}.

\subsubsection{Bifurcation scenario}\label{sec433}

\begin{figure}
\begin{center}
\includegraphics[width=0.48\textwidth]{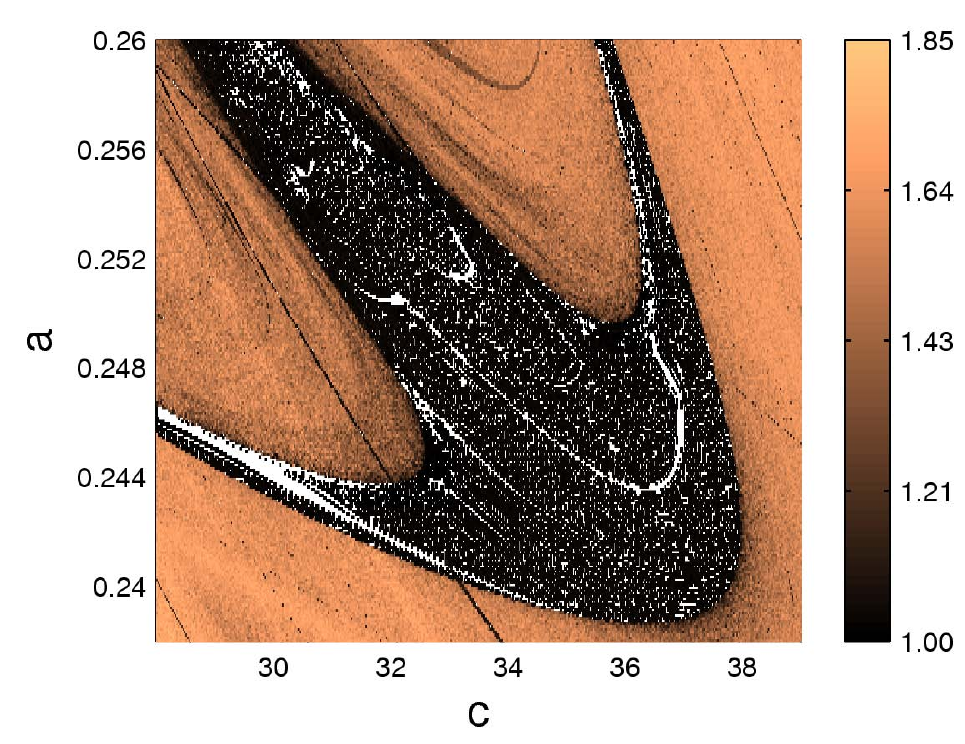}
\end{center}
\caption{Variation of the scale-local transitivity dimension $\hat{D}_{\mathcal{T}}(\varepsilon)$ obtained for individual realisations of the R\"ossler system with $N=5,000$ and $\rho=0.02$ in a two-dimensional cross-section $(a=b)$ of the parameter space. Periodic windows are characterised by minima of $\hat{D}_{\mathcal{T}}(\varepsilon)$ with values close to 1. White points indicate parameter combinations for which the numerical algorithms did not provide feasible results for the considered parameters (e.g., indicated an artificial fixed point behaviour due to the improper choice of the sampling rate).}
\label{fig:shrimp}
\end{figure}

The bifurcation scenario of the R\"ossler system is very rich and shows multiple complex bifurcations between periodic and chaotic solutions in dependence on its three control parameters. Recently, much interest has been spent on the investigation of so-called shrimps~\cite{Gallas1,Gallas2}, i.e., specific self-similar periodic windows with a complex shape that appear in certain two-dimensional subspaces of the full parameter space (see Fig.~\ref{fig:shrimp})~\cite{MTthesis,Bonatto2008PTRS,Gallas2010IJBC}. It has been demonstrated that statistical measures ba\-sed on recurrence quantification analysis as well as $\varepsilon$-re\-cur\-ren\-ce networks are well suited for automatically discriminating between periodic and chaotic dynamics and, hence, uncover complex bifurcations between both types of behaviour~\cite{Zou2010}. Within the complex network approach, transitivity properties have been found to be among the most suitable candidate measures for this purpose. Given the framework of our considerations presented in this work, this effect can be theoretically understood since periodic trajectories correspond to a lower-dimensional dynamics than chaotic ones, which is naturally detected by the transitivity dimension. 

\section{Summary}

The recently introduced $\varepsilon$-recurrence networks have a great potential for detecting qualitative changes in the dynamics of complex systems, which may correspond to nonstationarities, bifurcations, or different local attractor properties. While previous results have been mainly obtained numerically, this paper provides a theoretical fra\-me\-work for better understanding the links between network and attractor properties. In particular, we ha\-ve studied the local and global transitivity properties of $\varepsilon$-recurrence networks, which are closely interrelated with the local and global dimensionality of the studied attractor. This relationship motivated the definition of novel measures of dimensionality, the (local) clustering and (global) transitivity dimensions, which can be directly estimated from this type of networks. In this spirit, our corresponding results demonstrate that $\varepsilon$-recurrence networks provide an important link bet\-ween dynamical systems theory on the one hand, and graph theory on the other hand.

{We emphasise that many other established measures of dimensionality, such as box-counting and R\'en\-yi dimensions, are based on the scaling of local residence probabilities of typical trajectories on the attractor in different parts of the phase space with successively refined spatial resolution. As an exception, the correlation dimension is based on spatial two-point correlations in phase space. In this respect, clustering and transitivity dimensions are statistical properties of higher order, since they are based on geometric three-point interdependences, i.e., the mutual proximity of triples of state vectors on the attractor. As a result, local clustering dimensions allow quantifying the effective (possibly non-integer) local dimensions of the attractor in different parts of phase space. The fundamental importance of the corresponding geometric interpretation becomes visible in the representation of distinct spatial structures related with supertrack functions and UPOs, which cannot be detected by other traditional measures of dimensionality.

Beyond the aforementioned conceptual differences, we} note that our novel dimension measures have {further} important advantages in comparison to more traditional properties such as correlation or pointwise dimensions. These advantages mainly reflect the issue of practical estimation: whereas for many classical dimension measures, scaling properties of so\-me quantity have to be carefully evaluated (which requires large data sets and sophisticated estimation stra\-te\-gies~\cite{Sprott2003}), there is no need for considering any specific scaling for estimating clustering and transitivity dimensions. Besides the fact that the estimation becomes more direct, this also allows numerically obtaining reasonable estimates from rather short time series (i.e., data sets of size $N\sim {\cal O}(10^3\dots 10^4)$) at least for low-dimensional systems. {The required amount of data is therefore significantly lower than for classical properties such as $D_2$, implying that all numerical calculations performed for this paper can be completed on standard desktop computers within a reasonable amount of time. We expect that this advantage of much lower requirements with respect to the number of data should persist for higher-dimensional systems.} 

In contrast to these benefits, we have identified situations where our new measures behave pathologically (e.g., exceed the non-fractal dimension of the phase space in which the attractor is embedded). However, we emphasise that similar pathologies may also be found for other concepts of fractal dimension, e.g., due to the breakdown of the supposed scaling relationships, or in terms of the ``artificial'' upper bound of the Lyapunov dimension. The numerical examples discussed in this paper demonstrate that there is no simple relationship with any existing dimension measure, i.e., clustering and transitivity dimensions do not serve as bounds to any of the more traditional concepts, but typically have values that are comparable with those of other types of fractal dimensions estimated from the same trajectories.

Our theoretical considerations also confirm recent numerical results on the relationship between local transitivity properties and the location of dynamically invariant objects. Specifically, for the logistic map, high values of the local clustering coefficient coincide with the positions of supertrack functions, which has been studied in more detail in this work. For the three-di\-men\-sio\-nal chaotic R\"ossler oscillator, it has been shown that unstable periodic orbits with low periods coincide with local maxima of the same vertex property~\cite{Donner2010NJP}. Our results suggest that these findings can be generalised to other (discrete as well as time-continuous) complex systems, given that the invariant density of the attractor is sufficiently continuous in phase space. Examples such as Cantor sets or the two-dimensional H\'enon map have been discussed as well, illustrating the fact that in particular the proper estimation of local (pointwise) dimension measures is non-trivial for attractors with a pronounced fractal structure. {Our findings suggest a fundamental relationship between the differences of upper and lower clustering/transitivity dimensions (which have been found for certain self-similar sets) on the one hand, and the smoothness properties of the attractor on the other hand. A more detailed investigation of the corresponding interdependences will be subject of future studies.}

The relationship between local transitivity properties and local attractor geometry theoretically justified in this paper has some important consequences for possible practical applications of $\varepsilon$-recurrence networks in dynamical systems research. In particular, the fact that the local clustering dimensions are excellent candidates for quantitatively characterising the (mean) dimensionality of the system with\-in some $\varepsilon$-ball around any specific point on the attractor can help numerically identifying dynamically invariant objects such as unstable periodic orbits (or invariant manifolds of hyperbolic fixed points), which is still a problem of intensive scientific research~\cite{Saiki2007}. As a consequence, we emphasise that our transitivity-based dimension concept offer a novel approach for studying structures in the phase space of complex systems and appear to have meaningful and potentially relevant applications in both dynamical systems theory and real-world time series analysis.

\section*{Acknowledgements}

This work has been financially supported by the Leibniz association (project ECONS) and the Federal Ministry for Education and Research (BMBF) via the Potsdam Research Cluster for Georisk Analysis, Environmental Change and Sustainability (PROGRESS). {JFD acknowledges financial support by the German National Academic Foundation.} For calculations of complex network measures, the software package \texttt{igraph}~\cite{Csardi2006} has been used. Parts of the numerical calculations described in this work have been made using the IBM iDataPlex Cluster at the Potsdam Institute for Climate Impact Research.

\bibliographystyle{epj}
\bibliography{donner_epjb}

\end{document}